\documentclass[aps,preprintnumbers,prd,twocolumn,superscriptaddress, nofootinbib]{revtex4}
\usepackage{graphicx}
\usepackage{dcolumn}
\usepackage{bm}
\usepackage{hyperref}
\usepackage{epstopdf}
\usepackage{color}
\usepackage{hyperref}
\usepackage{epsfig,latexsym,cancel,amssymb,amsmath}
\graphicspath{{./Figures/}}

\def\lsim{\raisebox{-4pt}{$\,\stackrel{\textstyle{<}}{\sim}\,$}}
\def\gsim{\raisebox{-4pt}{$\,\stackrel{\textstyle{>}}{\sim}\,$}}

\def\beq{\begin{equation}} 
\def\eeq{\end{equation}} 
\def\bea{\begin{eqnarray}} 
\def\eea{\end{eqnarray}} 
\def\ben{\begin{enumerate}} 
\def\een{\end{enumerate}}

\def\lsim{\mathrel{\raise.3ex\hbox{$<$\kern-.75em\lower1ex\hbox{$\sim$}}}} 
\def\gsim{\mathrel{\raise.3ex\hbox{$>$\kern-.75em\lower1ex\hbox{$\sim$}}}} 
\def\ifmath#1{\relax\ifmmode #1\else $#1$\fi}

\newcommand{\met}{\mbox{${\not\! E}_{\rm T}$}}

\begin{document}
\DeclareGraphicsExtensions{.jpg,.pdf,.mps,.png}

\preprint{FERMILAB-PUB-14-183-T}
\preprint{LA-UR-13-24173}
\title{Supersymmetric Sub-Electroweak Scale Dark Matter, \\  the Galactic Center Gamma-ray Excess, \\ and Exotic Decays of the 125 GeV Higgs Boson }

\author{Jinrui Huang}
\email{jinruih@lanl.gov}
\affiliation{Theoretical Division, T-2, MS B285,
Los Alamos National Laboratory, Los Alamos, NM 87545, USA}

\author{Tao Liu}
\email{taoliu@ust.hk}
\affiliation{Department of Physics, The Hong Kong University of
  Science and Technology, Clear Water Bay, Kowloon, Hong Kong}

\author{Lian-Tao Wang}
\email{liantaow@uchicago.edu}
\affiliation{Enrico Fermi Institute, 
University of Chicago, Chicago, IL 60637, USA}
\affiliation{KICP and Dept. of Physics, University of Chicago, 5640
  S. Ellis Ave., Chicago, IL 60637, USA}

\author{Felix Yu}
\email{felixyu@fnal.gov}
\affiliation{Theoretical Physics Department, Fermi National
  Accelerator Laboratory, P.~O.~Box 500, Batavia, IL 60510, USA}

\today

\begin{abstract}

We continue our exploration of the nearly Peccei-Quinn symmetric limit
shared by common singlet extensions of the Minimal Supersymmetric
Standard Model.  This limit has been established as a viable framework
for studying sub-electroweak scale dark matter phenomenology and has
interesting and direct connections to new exotic Higgs decay physics.
We present analytic calculations to motivate the important
phenomenological features mentioned above.  We also discuss benchmark
points in this model framework that accommodate the observed Galactic
Center gamma ray excess.  We emphasize connections between
phenomenology of dark matter direct detection and indirect detection,
and new exotic decay channels for the 125 GeV Higgs boson.  We
conclude by identifying two benchmark modes of exotic Higgs decays for
$h \to \tau^+ \tau^- \met$ and $h \to b \bar{b} \met$ final states and
estimate their sensitivity prospects at the LHC.

\end{abstract}

\maketitle

\section{Introduction}
\label{sec:introduction}

As presented in two Letters~\cite{Draper:2010ew, Huang:2013ima}, the
Peccei-Quinn symmetry limit of singlet extensions of the Minimal
Supersymmetric Standard Model (MSSM) encompasses rich Higgs and dark
matter (DM) physics, whose phenomenology and collider signatures we
began exploring in~\cite{Huang:2013ima}.  The singlet extensions, like
the next-to-MSSM (NMSSM)~\cite{NMSSM}, nearly-MSSM
(nMSSM)~\cite{nMSSM}, and $\mu\nu$SSM~\cite{Fidalgo:2011ky}, are
traditionally motivated as possible solutions to the notorious $\mu$
problem in the MSSM and give rise to some limited scenarios for exotic
Higgs decays and DM phenomenology.  It was remarked, however, that
these scenarios share a Peccei-Quinn (PQ) symmetry limit ({\it cf.}
Ref.~\cite{Barger:2006dh}), giving rise to novel Higgs and DM
signatures outside the scope of previous studies.

Most notably, the PQ symmetry limit provides a supersymmetric
framework for studying sub-electroweak (sub-EW) scale
DM~\cite{Draper:2010ew}.  Possibilities for sub-EW scale DM in typical
MSSM contexts were stymied by LEP constraints and acceptable relic
density requirements~\cite{Arbey:2012na}.  Having a sub-EW DM
particle, however, neatly dovetails with exciting opportunities in
Higgs physics, as the 125 GeV Higgs discovered by CMS~\cite{:2012gu}
and ATLAS~\cite{:2012gk} has new exotic decay channels available.
This connection between DM physics and exotic Higgs decays in the PQ
symmetry limit scenario was explicitly explored in
Ref.~\cite{Huang:2013ima} for two benchmark models that gave rise to
two distinct exotic Higgs signatures of $\mu^+ \mu^- \met$ and $b
\bar{b} \met$ from the 125 GeV Higgs decay.  In this work, we will
continue exploring this connection between DM physics and Higgs
collider phenomenology with two new benchmarks.  The first, which will
give the exotic Higgs decay channel $\tau^+ \tau^- \met$, will help
complete the set of possible observable or theoretically
well-motivated decay modes for the 125 Higgs boson.  The second will
be a new benchmark corresponding to the $b \bar{b} \met$ exotic decay,
which instead of being optimized for discovery at the LHC, would
instead be motivated as a possible model explaining the Galatic Center
gamma ray excess~\cite{Goodenough:2009gk, Hooper:2010mq,
  Hooper:2011ti, Abazajian:2012pn, Hooper:2013nhl, Gordon:2013vta,
  Abazajian:2014fta, Daylan:2014rsa}.  Other discussions of models and
associated phenomenology for the Galactic Center (GC) gamma ray excess
include Refs.~\cite{Barger:2010mc, Ng:2013xha, Okada:2013bna,
  Modak:2013jya, Boehm:2014hva, Hardy:2014dea, Finkbeiner:2014sja,
  Alves:2014yha, Berlin:2014tja, Agrawal:2014una, Izaguirre:2014vva,
  Cerdeno:2014cda, Ipek:2014gua, Kong:2014haa, Ko:2014gha,
  Boehm:2014bia, Abdullah:2014lla, Ghosh:2014pwa, Martin:2014sxa,
  Berlin:2014pya, Basak:2014sza, Cline:2014dwa, Detmold:2014qqa,
  Wang:2014elb, Fields:2014pia, Arina:2014yna, McDermott:2014rqa},
although some interesting non-DM explanations are discussed in
Refs.~\cite{Carlson:2014cwa, Bringmann:2014lpa}.

We will first review the phenomenology of the PQ-axion supermultiplet
and the nearly PQ-symmetry limit of the NMSSM and nMSSM in
Sec.~\ref{sec:theory}, which includes the sub-EW scale dark matter
physics in this scenario, with bounds from both direct detection
experiments and the parameter space favored by the GC gamma ray excess
covered, as well as its potential connection to exotic decays of the
125 GeV Higgs boson.  In Sec.~\ref{sec:LHC}, we detail our LHC search
strategies for the separate channels of $h_2 \rightarrow \tau^+ \tau^-
+ \met$ and $h_2 \rightarrow b \bar{b} + \met$.  We conclude in
Sec.~\ref{sec:conclusion}.  Detailed calculations of the mass
eigenvalues and eigenstates of the $CP$-even Higgs sectors in the
PQ-symmetry limit as well as the coupling $y_{h_2 a_1 a_1}$ are given
in Apps.~\ref{app:gauge} and~\ref{app:goldstone}.

\section{Phenomenology of the Peccei-Quinn Symmetry Limit}
\label{sec:theory}

As mentioned in Sec.~\ref{sec:introduction}, the PQ symmetry limit of
singlet extensions of the MSSM provides a supersymmetric context for
studies of sub-EW scale DM and the corresponding exotic Higgs decays.
In the PQ symmetry limit, the superpotential and soft
supersymmetry-breaking terms are given by
\begin{align}
\mathbf{W} &= \lambda \mathbf{S} \mathbf{H_u} \mathbf{H_d} \ , 
\nonumber \\
V_{\text{soft}} &= {m^2_{H_d}} |H_d|^2 + {m^2_{H_u}} |H_u|^2 + {m^2_S}|S|^2 
\nonumber \\
 &- (\lambda A_{\lambda} H_u H_d S + \text{ h.c.} ) \ ,
\label{eqn:PQlimit}
\end{align}
where $H_d$, $H_u$ and $S$ denote the neutral Higgs fields of the
${\bf H_d}$, ${\bf H_u}$ and ${\bf S}$ supermultiplets, respectively.
We will temporarily ignore any possible explicit breaking of the PQ
symmetry in Eq.~\ref{eqn:PQlimit}, {\it e.g.} an NMSSM superpotential
term $\kappa {\bf S}^3$, but we remark that such small terms are
typically required in more realistic scenarios.  In addition, we will
further narrow our focus to the decoupling limit given by $\lambda =
\frac{\mu }{ v_S} \lesssim \mathcal{O}(0.1-0.3)$ ($\langle S \rangle =
v_S$), which will simplify our analytic analysis as well as help avoid
a Landau pole problem.  
We will also assume that there is no explicit $CP$-violation in the
Higgs sector, since the current experimental data constrains large
mixing between the $CP$-even and $CP$-odd Higgs
states~\cite{Djouadi:2013qya, Shu:2013uua}.

In this scenario, there are three important characteristics
distinguishing it from typical MSSM or NMSSM-like scenarios, and we
will discuss each in turn.

\subsection{The Peccei-Quinn axion supermultiplet}
\label{subsec:axion}

First, because of the PQ-symmetry and supersymmetry (SUSY), there
simultaneously co-exist three particles of sub-EW scale: the gauge
singlet-like $CP$-even and $CP$-odd Higgs bosons, $h_1$ and $a_1$, and
the singlino-like neutralino, $\chi_1$, with their masses much lighter
than the scale of the PQ symmetry breaking and their phenomenology
approximately model independent.  These particles or mass eigenstates
are strictly reduced to the PQ axion supermultiplet (saxion ($s$),
axion($a$), and axino ($\tilde a$)) in the SUSY limit.

The Goldstone (or PQ axion) supermultiplet is represented by
\begin{equation}
{\bf A} = A + \sqrt{2} \theta \tilde{a} + \theta^2 F_A \ , \quad
A = \frac{1}{\sqrt{2}} (s + i a)
\end{equation}
after the PQ symmetry is spontaneously broken, where
\begin{equation}
{\bf A} = \sum_i \frac{q_i v_i}{v_{\rm PQ}}({\bf \Sigma_i} - v_i) \ ,
\label{eqn:axion1}
\end{equation}
in which $i = 1$, $2$, $3$, $v_i = \{ v_S, v_u, v_d \}$ and ${\bf
  \Sigma_i} = \{ {\bf S}$, ${\bf H_u}$, ${\bf H_d} \}$ are the
superfields charged under the PQ symmetry with $\langle H_u \rangle =
v_u$, $\langle H_d \rangle = v_d$ and $v = \sqrt{v_u^2 + v_d^2} = 174$
GeV.  Here $v_{\rm PQ} = \sqrt{\sum_i q_i^2 v_i^2}$ is the $U(1)_{\rm
  PQ}$ breaking scale and $q_i$ is an effective $U(1)_{\rm PQ}$ charge
of ${\bf \Sigma_i}$.  In our model, 
\begin{equation}
q_d = -q_S \sin^2 \beta \ , q_u = -q_S \cos^2 \beta \ ,
\label{eqn:axion2}
\end{equation}
and 
\begin{equation}
v_{\rm PQ} = |q_S| \sqrt{v_S^2 + \frac{\sin^2 2\beta}{4} v^2} \ ,
\label{eqn:axion3}
\end{equation}
where $\tan \beta = v_u / v_d$.  If ${\bf \Sigma_i}$ is not charged
under any other continuous symmetries, then $q_i$ is simply its
$U(1)_{\rm PQ}$ charge~\cite{Tamvakis:1982mw}.  The axion ($a$) mass
is protected by the Goldstone theorem, and it is related to the saxion
($s$) and axino ($\tilde{a}$) masses by SUSY.  If the $U(1)_{\rm PQ}$
symmetry is global and SUSY is unbroken, then we have $m_s = m_{\tilde
  a} = m_a=0$.  If SUSY is broken (this is often true when $v_i \neq
0$), the saxion and axino become massive while the axion remains
massless.  There are two sources which may contribute to the mass
splitting between the superfield components.  The first arise as
diagonal corrections in superspace.  The second is the separate real
scalar and fermion mixing with other massive particles.  In the latter
case, the mass eigenstates become misaligned with the original saxion
and axino states. For $\lambda \lesssim 0.1$, $v_S$ is of TeV scale or
above, since an EW scale $\mu$ is required by electroweak symmetry
breaking (EWSB).  This immediately leads to $v_{\rm PQ} \sim v_S \gg
v$.  The PQ symmetry breaking is then mainly controlled by the singlet
superfield ${\bf S}$, and ${\bf A}$ is hence ${\bf S}$-like.

The diagonal axino mass at tree level is given by
\begin{equation}
m_{\tilde a} = \frac{-\sum_i q_i^2 v_i F_i}{v_{\rm PQ}^2} ,
\end{equation}
where $F_i$ is the $F$-component of the $i^{\text{th}}$ chiral
supermultiplet which is charged under the $U(1)_{\rm PQ}$ symmetry.
Note $m_{\tilde a}$ falls to zero when none of the PQ-charged
$F$-terms obtain a vacuum expectation
value~\cite{Tamvakis:1982mw}. Given
\begin{eqnarray}
F_{\mathbf{H_d}} &=& \lambda v_S v_u \ , \nonumber \\
F_{\mathbf{H_u}} &=& \lambda v_S v_d \ , \nonumber \\
F_{\mathbf{S}} &=& \lambda v_d v_u \ ,
\end{eqnarray}
we have
\begin{equation}
m_{\tilde a} = -\frac{\lambda^2 v^2 \sin 2\beta}{\mu} + 
\sum\limits_i \mathcal{O} 
\left(\frac{\lambda^{5-i}}{ \tan^i \beta} \right) \ .
\end{equation}
Since $v_{\rm PQ} \approx v_S \gg v$, the axino is mostly singlino in
our case.  The contribution to the axino mass from mixing with the
gauginos is further suppressed by the product of $ \lambda v /\mu
\approx v / v_{\rm PQ}$ and a gauge coupling factor.  An axino as
light as $\mathcal{O}(10)$ GeV or even lighter is therefore quite
natural in this context.  We denote the axino and other neutralinos as
$\chi_i$, with $\chi_1$ (mostly singlino) being the lightest.  The
largest non-singlino content of $\chi_1$ is Higgsino, which is of the
order $\lambda v / \mu$ and hence can be very small in our case.  One
important constraint on this scenario is the contribution of $Z \to
\chi_1 \chi_1$ to the $Z$ invisible decay width. In our case, the $Z
\chi_1 \chi_1$ coupling is suppressed by $(\lambda v / \mu)^2$, which
is small in the decoupling limit since $\lambda \lesssim 0.1$ and $\mu
\sim v$~\cite{Draper:2010ew}.  At the same time, $Z \to \chi_1 \chi_2$,
if kinematically allowed, is only suppressed by $\lambda v / \mu$ and
can be more constraining.

Next we consider the saxion mass, which has been discussed in detail
in Refs.~\cite{Miller:2005qua, Draper:2010ew}. We briefly summarize
the main conclusions here.  At tree level, in the large $\tan \beta$
limit, the minimum of the scalar potential in our scenario satisfies
\begin{equation}
A_\lambda \approx \mu \tan \beta \gg m_Z. 
\label{eq:tree_relation}
\end{equation}
For later convenience, we introduce two parameters
\begin{equation}
\varepsilon' = \frac{A_\lambda}{\mu \tan \beta } - 1 \ , \quad
\varepsilon = \frac{\lambda \mu}{m_Z} \varepsilon' \ ,
\end{equation}
which characterize the deviation from the exact relation in
Eq.~(\ref{eq:tree_relation}).  After EWSB, the saxion mixes with two
$CP$ even Higgses.  We denote the lightest scalar as $h_1$. At tree
level, we have
\bea
(m_{h_1}^2)_{\rm tree}  &=&  - 4 v^2 \varepsilon^2 + 16 \frac{v^4}{m_Z^2} 
\varepsilon^4 \nonumber \\ 
&+& \frac{4 \lambda^2 v^2  }{ \tan^2 \beta} 
\left (1- \frac{ \varepsilon m_Z}{\lambda \mu} \right) 
\left (1+ \frac{ 2 \varepsilon \mu}{\lambda m_Z} \right) \nonumber \\
&+& \sum\limits_i \mathcal{O} \left( \frac{ \lambda^{5-i}}{\tan^i \beta} 
\right) \ .
\eea
We see that in the decoupling limit, $\lambda \ll 1$, $m_{h_1}$ can be
much smaller than the EW scale without too much fine tuning.  At the
same time, to avoid a tachyonic mass for $h_1$, there is an upper
limit
\bea
\varepsilon^2 < \frac{\lambda^2}{\tan^2 \beta} \ . 
\eea

Based on these discussions, we would expect a natural co-existence of
three light singlet or singlino-like particles, $h_1$, $a_1$, and
$\chi_1$, in the PQ symmetry plus decoupling limit of singlet
extensions of the MSSM.  We also expect some small explicit PQ
symmetry breaking which would generate a small mass for the axion,
$a_1$.  This feature is clearly shown in the NMSSM context
in~\cite{Draper:2010ew}, while the lightness of $\chi_1$ in the PQ
limit of the other MSSM-extensions was also noticed in
Refs.~\cite{Menon:2004wv, Barger:2005hb}.  The light singlino-like
$\chi_1$ provides a natural supersymmetric sub-EW scale DM candidate.
In particular, the existence of light $a_1$ and $h_1$ states allow
$\chi_1$ to achieve simultaneously the correct relic density and a
spin-independent direct detection cross section varying over several
orders, which is the focus of Subsec.~\ref{subsec:singlino}.  We
remark that this feature is absent in the $R$-symmetry limit of the
NMSSM, where the coupling of the cubic term in superpotential $\kappa$
is large, leading to large contributions to the masses of both
$m_{h_1}$ and $m_{\chi_1}$ at tree level.

\subsection{Sub-EW scale singlino DM}
\label{subsec:singlino}

Second, if we assume $R$-parity conservation, $\chi_1$ can be a good
DM candidate at the sub-EW scale, in contrast with usual MSSM
constructions.  The direct detection cross section for $\chi_1$ varies
within a large range and arises dominantly via the exchange of a light
scalar with nucleons.  Moreover, the relic density is driven by pair
annihilation of singlino-like neutralinos probing the light
pseudoscalar resonance~\cite{Draper:2010ew} and can match current
measurements.  While the correct DM relic density can also be achieved
via an $s$-channel exchange of the singlet-like CP-even Higgs boson,
this process has no $s$-wave contribution and suffers a $p$-wave
suppression: realistic scenarios typically require more fine-tuning,
and the variability of the direct detection cross section is much more
constrained for a given $m_{\chi_1}$.

In the past decade, motivated by a series of interesting direct
detection experimental results (see~\cite{Agnese:2013jaa,
  Angle:2011th, Aprile:2012nq, Akerib:2013tjd, Aalseth:2012if,
  Bernabei:2010mq, Angloher:2011uu, Agnese:2013rvf}), many studies of
sub-EW scale DM have been performed in various contexts (e.g.,
see~\cite{DMdirect,Draper:2010ew,Arbey:2012na}).  It was found,
however, that a strict MSSM context for sub-EW scale DM is not easy to
achieve.  Of the MSSM neutralinos, a wino-like or higgsino-like DM
candidate would have an associated chargino at about the same mass and
be ruled out from searches at LEP.  A bino-like DM candidate may be
feasible, but it potentially requires the existence of extra light
particles, such as sfermions, to mediate annihilation or
coannihilation for reducing the DM relic density to an acceptable
level, which has been generally excluded by SUSY searches at LEP and
the LHC.  These factors make sub-EW scale DM highly constrained in the
MSSM (for more discussions , e.g., see~\cite{Arbey:2012na}).  In the
singlet extensions of the MSSM, however, a light singlino-like
neutralino, which generally arises in the nearly PQ-symmetry limit, is
relatively unconstrained, and so the PQ-limit provides a
supersymmetric benchmark to explore sub-EW scale DM
phenomenology~\cite{Draper:2010ew} (for some of the subsequent studies
on DM physics in or related to this scenario, see~\cite{DLH}).

Sub-EW scale DM is now being revisited in the context of the GC gamma
ray excess~\cite{Goodenough:2009gk, Hooper:2010mq, Hooper:2011ti,
  Abazajian:2012pn, Hooper:2013nhl, Gordon:2013vta, Abazajian:2014fta,
  Daylan:2014rsa} interpreted as an indirect detection of
DM~\cite{Goodenough:2009gk, Barger:2010mc, Hooper:2011ti,
  Abazajian:2012pn, Gordon:2013vta, Abazajian:2014fta, Daylan:2014rsa,
  Ng:2013xha, Okada:2013bna, Modak:2013jya, Boehm:2014hva,
  Hardy:2014dea, Finkbeiner:2014sja, Alves:2014yha, Berlin:2014tja,
  Agrawal:2014una, Izaguirre:2014vva, Cerdeno:2014cda, Ipek:2014gua,
  Kong:2014haa, Ko:2014gha, Boehm:2014bia, Abdullah:2014lla,
  Ghosh:2014pwa, Martin:2014sxa, Berlin:2014pya, Basak:2014sza,
  Cline:2014dwa, Detmold:2014qqa, Wang:2014elb, Fields:2014pia,
  Arina:2014yna, McDermott:2014rqa}.  This gamma ray excess was
identified from data collected by the Fermi Gamma Ray Space Telescope,
and studies indicate the excess extends at least $10^\circ$ from the
GC, lessening the possibility of astrophysical fakes.  Fits to the
spectrum favor a roughly 30 GeV dark matter particle annihilating to
$b \bar{b}$, with an annihilation cross section corresponding to that
of a thermal relic, $\langle \sigma v_{\rm rel} \rangle = 3 \times
10^{-26}$ cm$^3$/s~\cite{Goodenough:2009gk, Barger:2010mc,
  Hooper:2011ti, Abazajian:2012pn, Gordon:2013vta, Abazajian:2014fta,
  Daylan:2014rsa}.  Again, this hint of sub-EW scale DM cannot be
accommodated in the MSSM, but we will show a suitable benchmark in the
PQ limit of singlet extensions of the MSSM.

In particular, the natural annihilation channel for 30 GeV singlinos
in the PQ symmetry limit is via $\chi_1 \chi_1 \to a_1 \to b \bar{b}$.
For $m_{\chi_1} \sim 30$ GeV and $m_{a_1} \sim 60$ GeV, the $a_1$
pseudoscalar preferentially decays to the heaviest kinematically open
pair of SM fermions, namely $b \bar{b}$.  For the GC gamma ray excess,
this will serve to avoid diluting the gamma ray spectrum shape.

The light $a_1$ and $h_1$ states are critical for helping $\chi_1$
achieve the appropriate relic density as well as determining the
direct detection possibilities.  Unlike an $s$-wave dominant
annihilation process, the thermally averaged annihilation cross
section $\langle \sigma v_{\rm rel} \rangle$ is sensitive to the
temperature for processes mediated by the Breit-Wigner enhancement
effect, {\it e.g.}, the one under consideration (for general
discussions on the Breit-Wigner mechanism in DM physics,
see~\cite{Griest:1990kh, Gondolo:1990dk}).  This is simply because the
chance for the annihilated DM particles to sit on or close to the
mediator resonance is temperature-dependent.  For example, if
$m_{a_1}$ is smaller than $2m_{\chi_1}$, as the temperature decreases,
the annihilated DM particles get less active and hence have a better
chance to sit close to the mediator resonance or have a larger
$\langle \sigma v_{\rm rel} \rangle$.  Explaining the GC gamma-rays
requires $\langle \sigma v_{\rm rel}\rangle \sim 10^{-26}~{\rm
  cm^3}/{\rm s}$. This means that $\langle \sigma v_{\rm rel}\rangle$
is smaller than this value in the early universe and the DM particles
therefore are generically over-produced.  So we will consider the case
$m_{a_1} > 2m_{\chi_1}$ only.  Different from the first case, $\langle
\sigma v_{\rm rel} \rangle$ tends to have a larger value in the early
Universe in this case ({\it e.g.}, see~\cite{Guo:2009aj}), leading to
\begin{equation}
\Omega h^2 \sim 10^{-4} \times \frac{0.5}{{\rm erfc} 
\left( \sqrt{\frac{m_{\chi_1}}{T_f} 
\left| 1- \frac{m_{a_1}^2}{4 m_{\chi_1}^2} \right| } \right)}
\label{eqn:relicdensity}
\end{equation}
To generate the observed DM relic density, therefore, new inputs like
non-thermal production mechanisms are needed. This can be achieved by
decaying thermally produced next-to-lightest supersymmetric particle
(NLSP) (with its density before the decay denoted by $\Omega_{\rm
  NLSP} h^2$), such as slepton, sneutrino or neutralino, with the DM
relic density given by
\begin{equation}
\Omega h^2 = \frac{m_{\chi_1}}{m_{\rm NLSP}} \Omega_{\rm NLSP} h^2.
\end{equation}
We will leave the detailed discussions of such mechanisms for future
work.

\subsection{Exotic Higgs decays}
\label{subsec:exotic}

The third main feature of the PQ symmetry limit is the set of new
exotic decays for the SM-like Higgs boson, which can be potentially
probed at colliders soon or in the future.  We first note that in the
PQ symmetry limit, unlike the $R$-symmetry limit, the exotic decay
channels of the SM-like Higgs into a pair of light singlet-like
$CP$-even or $CP$-odd Higgs bosons are generically suppressed.  The
suppressed tree-level couplings of the SM-like Higgs boson $h_2$ with
$h_1 h_1$ and $a_1 a_1$ can be directly calculated from the Higgs
potential, given by~\cite{Draper:2010ew}
\begin{eqnarray}
y_{h_2 a_1 a_1} &=& -\sqrt{2} \lambda\varepsilon \frac{m_Z v}{\mu}  + 
\sum\limits_i \mathcal{O} \left( \frac{\lambda^{4-i}}{\tan^i \beta} \right) 
\ , \nonumber \\ 
y_{h_2 h_1 h_1} &=& -\sqrt{2} \lambda \varepsilon \frac{m_Z v}{\mu}  + 
2 \sqrt{2}   \varepsilon^2 v \nonumber \\
&& + \sum\limits_i \mathcal{O} \left( 
\frac{\lambda^{4-i}}{\tan^i \beta} \right)
\label{eqn:yh2a1a1}
\end{eqnarray}
in the exact PQ limit.  Here we have used the mixing parameters given
in Eq.~(\ref{eqn:eigenstate}) in Appendix~\ref{app:gauge}.
Alternately, the coupling $y_{h_2 a_1 a_1}$ can be calculated using
the properties of the Goldstone boson, which we detail in
Appendix~\ref{app:goldstone}.  We see that both Br$(h_2 \to h_1 h_1)$
and Br$(h_2 \to a_1 a_1)$ are suppressed by $|\lambda \varepsilon|^2
\ll 1$.  Therefore, these decay channels are rather inconsequential
for the SM-like Higgs search in our scenario.  This is different from
the well-known physics in the $R$-symmetry limit of the
NMSSM~\cite{Dobrescu:2000jt}, where $a_1$ is light, playing a role of
pseudo-Goldstone boson in breaking the approximate $R$-symmetry.
There, the decay of the SM-like Higgs boson $h_2 \to a_1 a_1$ is
typically significant.

\begin{figure}[ht]
\begin{center}
\includegraphics[width=0.45\textwidth]{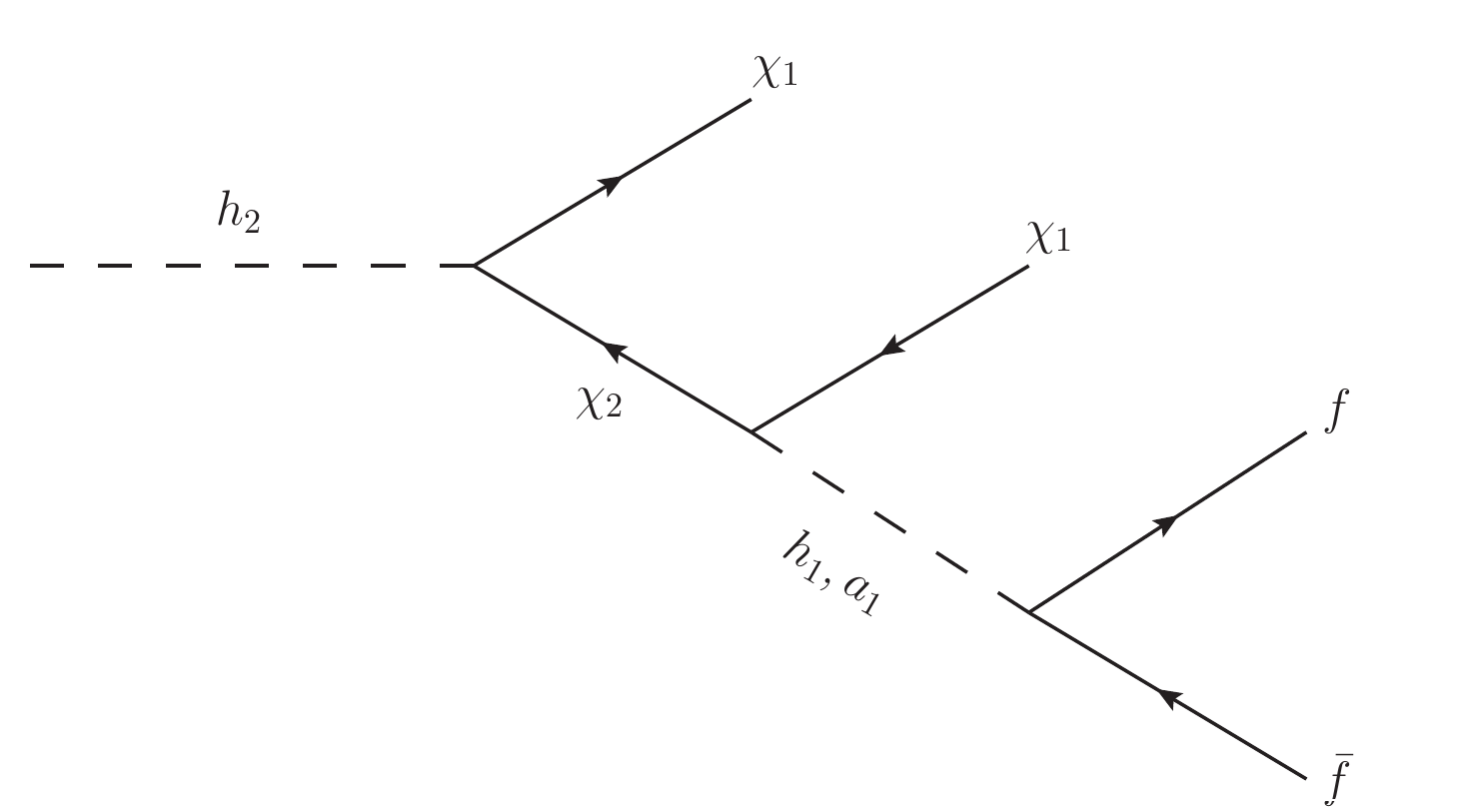}
\caption{A new decay channel for the SM-like Higgs boson, $h_2$, in
  the PQ-symmetry limit of the NMSSM.}
\label{fig:decay_topo}
\end{center}
\end{figure}

The suppression of the $h_2 \to h_1 h_1$, $a_1 a_1$ decay channels 
and the lightness of singlino-like $\chi_1$  in the PQ 
symmetry limit open up possibilities for a new category of exotic 
decays of the 125 GeV Higgs boson which are initiated by decay 
into two fermions~\cite{Huang:2013ima}:   
\begin{itemize}
\item if $\chi_2$ is bino-like and satisfies $m_{\chi_2} < m_{h_2} -
  m_{\chi_1}$, then $h_2$ can decay significantly via $h_2\to \chi_1
  \chi_2$;
\item if $\chi_2$ is bino-like and satisfies $m_{\chi_2} <
  \frac{m_{h_2}}{2}$, then $h_2$ can also decay significantly via
  $h_2\to \chi_2 \chi_2$, though its branching ratio is relatively
  small, compared with $h_2\to \chi_1 \chi_2$, due to the phase space
  suppression.
\end{itemize}
The bino-like neutralino subsequently decays via the following main
ways~\cite{Huang:2013ima}:
\begin{itemize}
\item $\chi_2 \rightarrow \chi_1 h_1/a_1 \rightarrow \chi_1 f
  \bar{f}$, which is favored most by kinematics in general context,
  due to the lightness of $h_1/a_1$.
\item $\chi_2 \rightarrow \chi_1 Z/Z^* \rightarrow \chi_1 f \bar{f}$,
  though this chain is strongly constrained by the availability of the
  phase space.
\item $\chi_2 \rightarrow \chi_1 \gamma$, whose branching ratio can be
  as large as $\mathcal O(0.01-0.1)$ in the case where the mass
  splitting between $\chi_1$ and $\chi_2$ is small, say,
  $m_{\chi_2}-m_{\chi_1} < m_{a_1/h_1}, m_Z$~\cite{Curtin:2013fra}.
\end{itemize}
The complete decay chain which is favored most by kinematics is shown
in Fig.~\ref{fig:decay_topo}\footnote{This decay topology was also
  noted in a simplified model context~\cite{Chang:2007de}.}. In
addition, though it is not the focus of this article, if
$m_{h_2}\lesssim m_{\chi_1} + m_{\chi_2}$, $h_2\to \chi_1\chi_1$ can
be significant.  So, the decay topologies of $h_2$ in this scenario
are very rich. As a matter of fact, for the seven possible topologies
listed in Fig.~2 of Ref.~\cite{Curtin:2013fra}, all of them can be
achieved in this scenario except that the one $h_2 \to 2 \to 4$ is
generically suppressed.

The richness of the $h_2$ decay topologies necessarily leads to
variety of its kinematics at colliders. However, all of these decay
chains eventually give a final state with at least one singlino-like
$\chi_1$, the collider signatures therefore are generically
semivisible (here we will not consider the possibility $h_2\to
\chi_1\chi_1$) or characterized by missing transverse energy (MET) and
some visible objects, with or without a resonance. For example, we
have
\begin{eqnarray}
h_2 \to \met + b\bar b, \ \tau^+ \tau^-, \ \ell^+ \ell^-,
\ \gamma\gamma, \ \gamma \ ,
\end{eqnarray}
for $h_2 \to \chi_1 \chi_2$. The visible objects can be either
collimated, via the decays of the light resonance $a_1/h_1$, or
isolated, via the decays of the $Z$ boson. The discussion can be
generalized to $h_2 \to \chi_2 \chi_2$ though its final state is more
complicated and the decay products tend to be softer. The sensitivity
of the LHC to these possibilities is mainly driven by the final state
signature. Here we will focus on the topology shown in
Fig.~\ref{fig:decay_topo}, considering its novelty and its favoredness
by kinematics.

For $h_1$, $a_1 \lesssim 1$ GeV, the corresponding decay $h_2 \to
\mu^+ \mu^- \met$ is easy to identify at the LHC with a specialized
muon-jet identification procedure, which we detailed
in~\cite{Huang:2013ima} and has been highlighted as one of the highly
motivated exotic Higgs searches within the 7 and 8 TeV data set of the
LHC in Ref.~\cite{Curtin:2013fra}.  For $1 \lesssim h_1$, $a_1
\lesssim 4$ GeV, the decay to strange mesons or a gluon pair is
typically dominant~\cite{Curtin:2013fra}. They are very difficult to
detect, and we do not expect that any sensitivity can be obtained with
conventional cut and count analyses.  But, potentially this parameter
region can be still probed by $h_2 \to \mu^+ \mu^- \met$ with a future
data set from the LHC run II program, given Br$(a_1\to \mu^+\mu^-) \sim
\mathcal O(0.01-0.1)$ for $\tan\beta > 1$~\cite{Curtin:2013fra}.

For $4 \lesssim h_1$, $a_1 \lesssim 10$ GeV, the decay to tau pairs and
missing transverse energy can be potentially probed with large amounts of
integrated luminosity, which we discuss in Subsec.~\ref{subsec:tata}.
For scalar and pseudoscalar masses larger than 10 GeV (and given the
decay topology in Fig.~\ref{fig:decay_topo}), the dominant decay to $b
\bar{b}$ is possible to probe at the LHC.  We studied one discovery
possibility in~\cite{Huang:2013ima}, but in Subsec.~\ref{subsec:bb},
we will focus on a new benchmark motivated by the GC gamma ray excess.

As a last comment in this section, we note that the monojet searches
at the LHC ({\it cf.}~\cite{Goodman:2010yf}) generally have no
sensitivity to this scenario, even if $\sigma_{\text{SI}}$ is as large
as $10^{-40}$ cm$^2$.  This is because the singlet-like mediator
typically has a small production cross section and in addition, mainly
decays into the SM fermions.

\begin{figure*}[t]
\begin{center}
\includegraphics[width=0.6\textwidth]{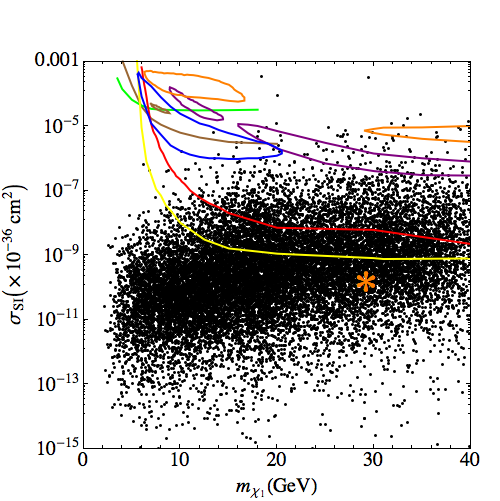}
\caption{Spin-independent direct detection cross section for $\chi_1$
  in the nearly PQ-symmetry limit of the NMSSM.  The scan is over all
  parameters, in the ranges $0.05 \leq \lambda \leq 0.3$, $0.0005 \leq
  \kappa \leq 0.03$, $|\varepsilon'| = \left| \frac{A_\lambda}{\mu
    \tan\beta} -1\right| \leq 0.25$, $-120 \leq A_\kappa \leq 0$ GeV,
  $5 \leq \tan \beta \leq 25$, and $100 \leq \mu \leq 400$ GeV.  We
  have assumed soft squark masses of 2 TeV, slepton masses of 200 GeV,
  $A_{u,d,e} = -3.5$ TeV, and bino, wino and gluino masses of
  $80-120$, 200, and 2000 GeV, respectively. The black points have a
  relic density $\Omega h^2 \leq 0.131$ (a default choice set in
  NMSSMTools 4.2.1~\cite{NMSSMTools}).  The curves show limits at
  90$\%$ C.L. from the CDMSlite~\cite{Agnese:2013jaa} (green), updated
  XENON10 S2-only~\cite{Angle:2011th} (brown),
  XENON100~\cite{Aprile:2012nq} (red), and LUX~\cite{Akerib:2013tjd}
  (yellow) analyses.  The contours identify possible signal regions
  associated with data from CoGeNT~\cite{Aalseth:2012if} (brown,
  90$\%$ C.L.), DAMA/LIBRA~\cite{Bernabei:2010mq} (orange, 99.7$\%$
  C.L.), CRESST~\cite{Angloher:2011uu} (purple, 95.45\% C.L.), and
  CDMS II Si~\cite{Agnese:2013rvf} (blue, 90$\%$ C.L) experiments.
  The orange star corresponds to the benchmark presented in
  Table~\ref{DMbm}.}
\label{fig:DM}
\end{center}
\vglue 0.4cm
\end{figure*}

\begin{figure*}[t]
\begin{center}
\includegraphics[width=0.3\textwidth]{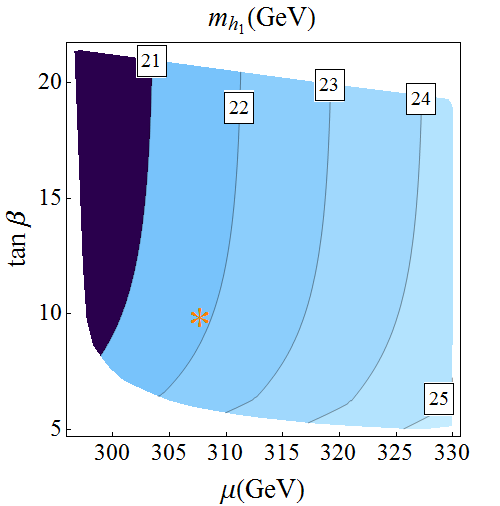}
\includegraphics[width=0.3\textwidth]{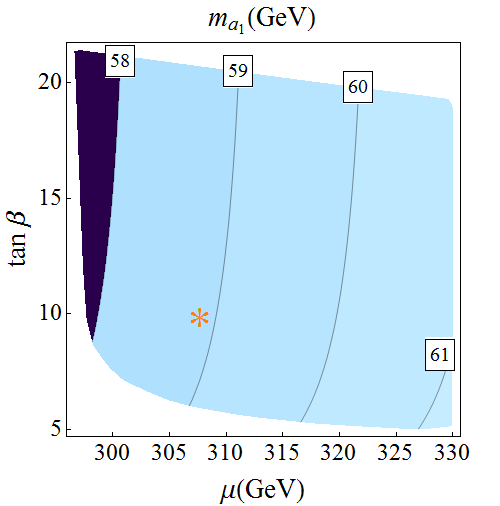}
\includegraphics[width=0.3\textwidth]{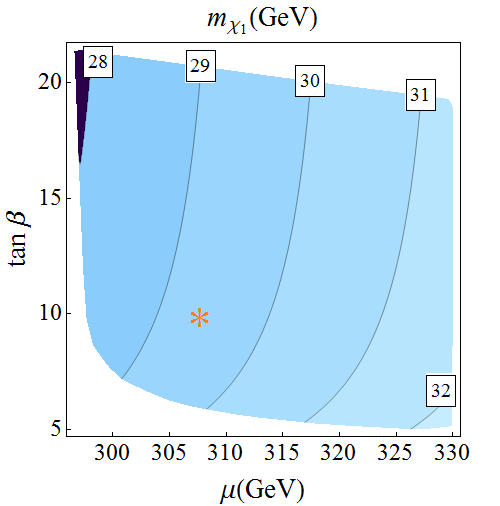}
\includegraphics[width=0.3\textwidth]{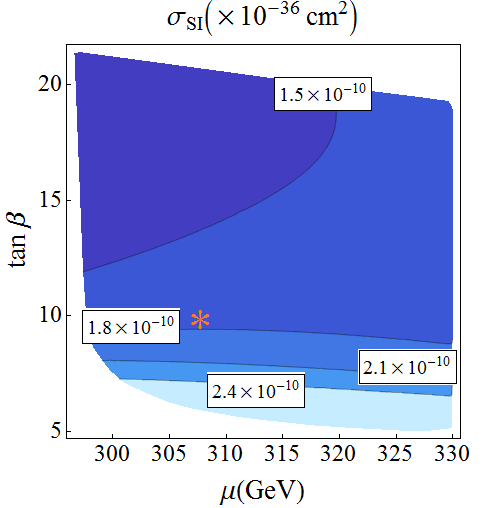}
\includegraphics[width=0.3\textwidth]{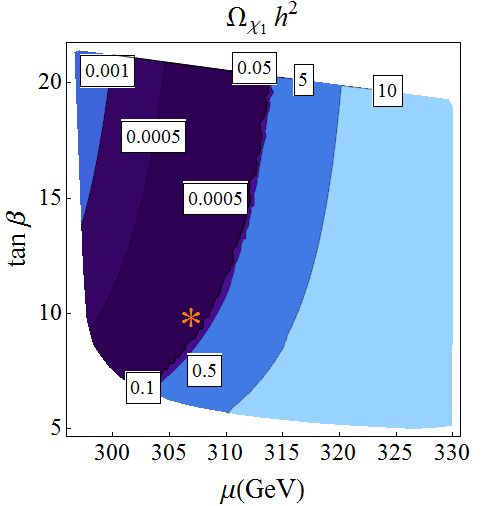}
\includegraphics[width=0.3\textwidth]{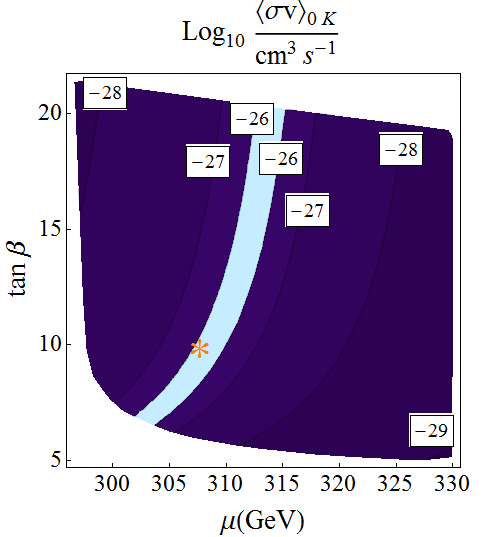}
\includegraphics[width=0.3\textwidth]{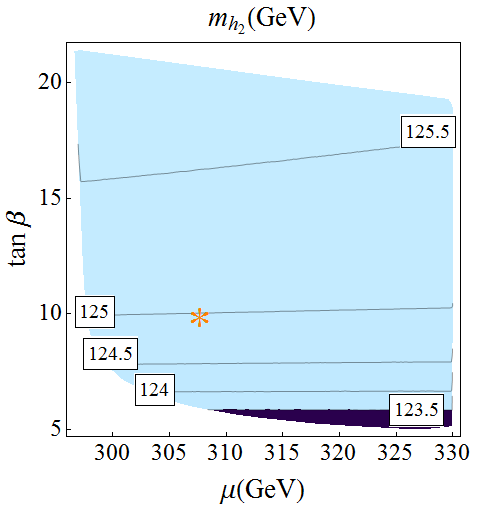}
\includegraphics[width=0.3\textwidth]{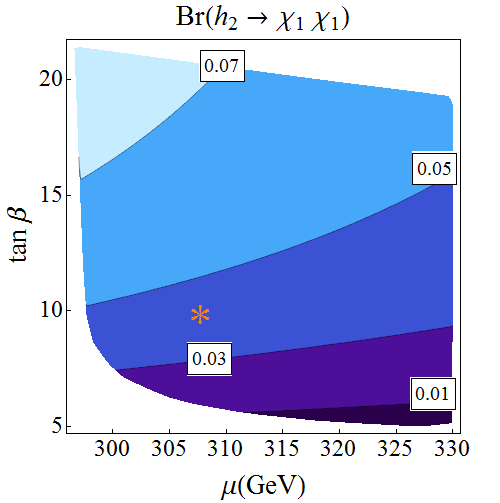}
\includegraphics[width=0.3\textwidth]{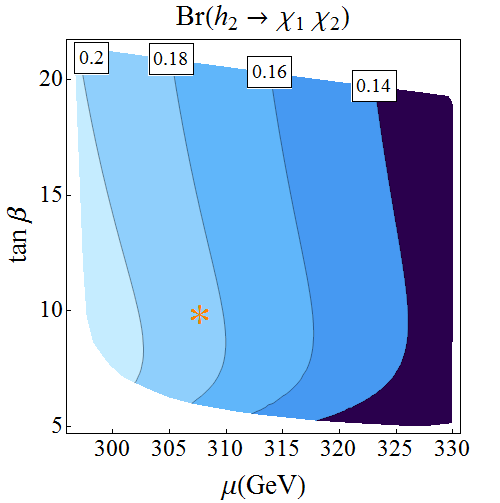}
\caption{Embedding the gamma ray excess from the GC as DM signals in
  the nearly PQ-symmetry limit of the NMSSM.  Here $\lambda = 0.21$,
  $\kappa \leq 0.01$, $\varepsilon' = - 0.01$, and $A_\kappa = -78$ GeV
  have been assumed. In addition, soft SUSY-breaking parameters for
  gauginos, squarks, and sleptons are set to be $M_1 = 85$ GeV, $M_2 =
  200$ GeV, $M_3=2000$ GeV, $\tilde M_l^2 = (200 {\rm GeV})^2$, and
  $\tilde M_q^2 = (2000 {\rm GeV})^2$, respectively; soft
  SUSY-breaking trilinear parameters are assumed to be
  $A_u=A_d=A_l=-3500$ GeV. The orange star represents the benchmark
  point shown in Table~\ref{DMbm}.}
\label{fig:DM2}
\end{center}
\vglue 0.4cm
\end{figure*}

\begin{table*}[t]
\begin{center}
\begin{tabular}{|c|c|c|c|c|c|c|c|c|}
\hline
 $\lambda$ & $\kappa$ & $A_\lambda$ 
& $A_\kappa$  & $\mu$  & $\tan \beta$ 
& $m_{h_1}$ & $m_{a_1}$ & $m_{\chi_1}$ \\
\hline
0.21 & 0.01& 3047.2  & $-78.0$ & 307.8 & 10.0 & 21.9  & 58.9  & 29.4   \\ \hline
$m_{h_2}$ & $m_{\chi_2}$ & Br$(h_2\to \text{SM})$ 
& Br$(h_2\to \chi_1\chi_2)$ & Br$(\chi_2 \to \chi_1 h_1)$ 
& Br$(h_1 \to b\bar b)$ &  $\Omega h^2 (10^{-5})$ & $\sigma_{\rm SI} (10^{-46})$ 
& $\langle \sigma v\rangle_{0K}(10^{-26})$ \\ \hline
125.0 & 80.9 & 75.0\% &  18.5\% & 100\% & 86.7\% &  
8.0 & 1.7 & 2.5 \\ \hline
\end{tabular}
\end{center}
\caption{Benchmark of sub-EW DM in the nearly PQ-symmetry limit of the
  NMSSM which is dedicated to encoding the event excess of cosmic
  gamma-ray from the GC.  Soft SUSY-breaking sfermion and
  gaugino parameters are as given in the caption of
  Fig.~\ref{fig:DM2}.  All mass parameters are in GeV, and
  $\sigma_{\rm SI} $ and $\langle \sigma v\rangle_{0K}$ have units of
  cm$^2$ and cm$^3/s$, respectively.
\label{DMbm}} 
\vglue 1.0cm
\end{table*}

\subsection{Parameter space scan}
\label{subsec:parameterspace}

We can illustrate many of the features described in
Subsecs.~\ref{subsec:axion}--~\ref{subsec:exotic} with a general
parameter space scan in the PQ symmetry limit of the NMSSM.

In Fig.~\ref{fig:DM}, we show the current bounds of DM direct
detections on this scenario, with the black points resulting in a
relic density $\Omega h^2 \leq 0.131$ and satisfying all of the other
built-in constraints in NMSSMTools 4.2.1~\cite{NMSSMTools}, such as
from Higgs searches, superpartner searches, muon $g-2$, flavor
physics, invisible $Z$-decay, and the constraints from $\Upsilon$
decays. The curves show limits at 90$\%$ C.L. from the
CDMSlite~\cite{Agnese:2013jaa} (green), updated XENON10
S2-only~\cite{Angle:2011th} (brown), XENON100~\cite{Aprile:2012nq}
(red), and LUX~\cite{Akerib:2013tjd} (yellow) analyses.  The contours
identify possible signal regions associated with data from
CoGeNT~\cite{Aalseth:2012if} (brown, 90$\%$ C.L.),
DAMA/LIBRA~\cite{Bernabei:2010mq} (orange, 99.7$\%$ C.L.),
CRESST~\cite{Angloher:2011uu} (purple, 95.45\% C.L.), and CDMS II
Si~\cite{Agnese:2013rvf} (blue, 90$\%$ C.L) experiments.  The orange
star corresponds to the benchmark presented in Table~\ref{DMbm}.  The
unconstrained parameter space in this scenario can be readily probed
by future iterations of current DM direct detection experiments.

In Fig.~\ref{fig:DM2}, we show physics in the neighborhood of the
benchmark point (orange star) which is used to explain the GC
gamma-ray excess. The panels in the top row show the simultaneous
smallness of $m_{h_1}$, $m_{a_1}$ and $m_{\chi_1}$.  The panels in the
middle row show the relic density given by a pair annihilation
enhanced by the $a_1$-mediated Breit-Wigner effect, as well as the
thermally averaged annihilation cross section in the Universe
nowadays. Indeed, one can find a narrow band where $\langle \sigma v
\rangle_{0K} \sim 10^{-26}$cm$^3$s$^{-1}$.  Yet, the overlap of this
band with the correct $\Omega_{\chi_1} h^2$ is a very small slice of
the parameter space.  Obtaining the correct relic abundance requires a
delicate fine-tuning of parameters if the annihilation mechanism
through the $a_1$ pseudoscalar is the only avenue available, and
indeed, the benchmark presented in Table~\ref{DMbm} does not have an
appropriate $\Omega_{\chi_1} h^2$ to saturate the DM relic density.
As emphasized before, however, non-thermal production mechanisms of
the DM can alleviate this problem and are, in fact, required for much
of the parameter space.  This can be satisfied by decays of thermally
produced NLSPs, which put nontrivial constraints on the remaining
sparticle spectrum outside of the exotic Higgs decay signatures.  The
cosmological constraints and limits from direct collider searches and
indirect flavor observables on the required sparticle spectrum to
non-thermally produce the $\chi_1$ DM is a discussion we will reserve
for future work.  Nonetheless, the panels in the last row indicate
that, if $\chi_2$ is bino-like and of the sub-EW scale, the decay of
$h_2 \to \chi_1\chi_2$ is significant.

\section{Exotic Higgs decays search strategies at the LHC}
\label{sec:LHC}

We now focus on the possibilities for exotic decays of the SM-like
Higgs boson that arise from the PQ-symmetry limit.  As depicted in
Fig.~\ref{fig:decay_topo}, we study the decay chain $h_2 \to \chi_1
\chi_2$, with $\chi_2 \to \chi_1 h_1$, $\chi_1 a_1$, and $h_1$, $a_1
\to f \bar{f}$. For our benchmark scenario, the collider signature
will be $f \bar{f} + \met + X$, where the SM fermions $f \bar{f}$ are
produced via the decay of the light resonance $h_1$ or $a_1$ and
appear as a pair of collimated leptons, jets, or merge into a fat jet
at the LHC, and $X$ denotes the decay products of the particles
produced in association with $h_2$.  We have studied the cases with $f
\bar{f} = \mu^+ \mu^-$, $b \bar{b}$ in Ref.~\cite{Huang:2013ima}.
Here, we will extend our analysis to include the $\tau^+ \tau^-$ final
state as well as consider a modified benchmark for the $b \bar{b}$
channel that is motivated by the connection to the GC gamma ray
excess.

Similar to Ref.~\cite{Huang:2013ima}, we introduce a scale factor
\begin{eqnarray}
c_{\text{eff}} & = & \frac{\sigma(p p \to h_2)}{\sigma(p p \to h_{\rm SM})} 
\times {\rm Br}(h_2 \to \chi_1 \chi_2) \\ \nonumber
& & \times {\rm Br}(\chi_2 \to h_1 \chi_1) 
\times {\rm Br}(h_1 \to f \bar{f}) \ ,
\label{eqn:ceff}
\end{eqnarray}
where $\sigma(p p \to h_2)$ and $\sigma(p p \to h_{\text{SM}})$ are
the production cross sections for the SM-like and SM Higgs boson (the
first calculated in the PQ-symmetry limit of the NMSSM and the second
calculated in the SM) in the relevant production mode, and we assume
the narrow width approximation for each intermediate decaying
particle.  The current limits on an invisible or unobserved decay
width for the Higgs boson are as large as 60\% (25\%) at the 95\%
C.L., with an enhanced $\Gamma(h_2\to gg)$ allowed (not allowed)
~\cite{ATLAS:2013sla}.  As the limit on a non-standard Higgs decay
width improves, our final sensitivity results can be rescaled by the
$c_{\rm eff}$ factor, in the same spirit as the simplified model
framework~\cite{Alves:2011wf}.  We also note that exotic production
modes for the SM-like Higgs boson could increase the effective
production rate~\cite{Yu:2014mda}.  The effective rate for our signal,
however, not only derives from direct exotic decays of the SM-like
Higgs but also from alternative production modes of the light NMSSM
resonances in our model.  From Table~\ref{DMbm}, we note that the
bino-like $\chi_2$ has a 100\% branching fraction to $\chi_1 h_1$:
this is reminiscent of gauge-mediation SUSY models with a light
gravitino, where every SUSY cascade in our model ends with a $\chi_2$
NLSP that subsequently decays to $\chi_1 h_1$.  Hence, the LHC
prospects of discovering the signature of light NMSSM resonances could
be greatly enhanced via non-Higgs exotic decays and not be constrained
by global fits to the invisible or exotic decay branching fraction of
the SM-like Higgs.  Although our analyses are optimized for finding
the light NMSSM resonances via their kinematics as decay products of
the SM-like Higgs, these additional modes would certainly improve the
sensitivity prospects as long as the new sparticles are within the
reach of the LHC.  Hence, the $c_{\text{eff}}$ scaling factor defined
in Eq.~(\ref{eqn:ceff}) only captures a piece of the potential signal
yield for $h_1 + \met$ production.  Clearly, though, an optimized
analysis of these additional modes would require a separate collider
analysis, which we reserve for future work.

In our analyses, the signal and background samples for both analyses
are simulated using MadGraph 5~\cite{Alwall:2011uj} with CTEQ6L1
parton distribution functions~\cite{Pumplin:2002vw} and MLM
matching~\cite{Mangano:2001xp, Mangano:2002ea}, with a matching scale
$Q=30$GeV.  The $h_2$ decays are handled in MadGraph 5 implementing an
NMSSM model file based on Ref.~\cite{Draper:2010ew}.  These events are
showered and hadronized using \textsc{Pythia}
v6.4.20~\cite{Sjostrand:2006za}.  Jet clustering is performed with the
\textsc{FastJet} v.3.0.2~\cite{Cacciari:2011ma} package.  As with
Ref.~\cite{Huang:2013ima}, we use a mock detector simulation
incorporating ATLAS and CMS performance results on
jets~\cite{Chatrchyan:2011ds}, electrons~\cite{Aad:2011mk},
muons~\cite{ATLAS:2011hga}, and missing
energy~\cite{Chatrchyan:2011tn}.

\subsection{Case I: $h_2 \to \tau^+\tau^- + \met$}
\label{subsec:tata}

For $4 \lesssim m_{h_1} \lesssim 10$ GeV, the dominant decay of $h_1$
proceeds via two tau leptons.  For concreteness, we adopt the
benchmark indicated in Table~\ref{table:benchmark1}.  Because of the
small $h_1$ mass, the two taus are relatively soft and only obtain
their boost from the kinematics of the cascade decay.  So we adopt the
SM Higgs production mode $Z h_2$ and we will trigger on $Z \to \ell^+
\ell^-$, $\ell = e$ or $\mu$.  The moderate boost to $h_2$ provided by
the recoiling $Z$ combined with the available phase space from the
cascade decay $h_2 \to \chi_2 \chi_1$, $\chi_2 \to \chi_1 h_1$, $h_1
\to \tau^+ \tau^-$ will serve to roughly collimate the ditau pair.  We
will focus on the tau decays characterized by one-prong and
three-prong tracks, where the prongs include both charged pions as
well as charged leptons.  The alternate high $p_T$, isolated leptonic
tau decays will likely be very difficult to identify because of the
loss of statistics and the characteristic softness of the leptons.
Then, the signal is characterized by an opposite-sign (OS), same-flavor (SF)
dilepton $Z$ candidate, a jet with track counts consistent with tau
parents, and missing transverse energy.

\begin{table} [ht]
\begin{tabular}{c|c|c|c|c}\hline\hline
                    & $m_{h_1}$ & $m_{h_2}$ & $m_{\chi_1}$ & $m_{\chi_2}$ \\ 
\hline
$h_1 \to \tau^+\tau^-$ & 8 GeV   & 125 GeV  & 10 GeV     & 80 GeV  \\ 
\hline
\end{tabular}
\caption{Benchmark used for the collider analysis of $h_2 \to
  \tau^+\tau^- + \met$.}
\label{table:benchmark1}
\end{table} 

The SM backgrounds for this challenging signal are $Z + $ jets, $Z \to
\ell^+ \ell^-$, fully leptonic $t \bar{t}$, fully leptonic $W^+ W^-$,
fully leptonic $W^\pm Z$, and $ZZ \to \ell \ell \nu \nu$.  For the
electroweak $Z + $ jets and the diboson backgrounds, we adopt a flat
$K$-factor of 1.3.  The $t \bar{t}$ background is normalized to 833 pb
at 14 TeV LHC~\cite{Bonciani:1998vc} to account for next-to-leading
order (NLO) + next-to-leading logarithm QCD corrections.  The signal
cross section for $Z h_2$ production was fixed to 0.9690 for $h_2$ 125
GeV~\cite{Heinemeyer:2013tqa}, which includes next-to-next-to-leading
order QCD + NLO EW corrections.  We adopt $c_{\text{eff}} = 1$, and our
results can be readily rescaled for other values.

For efficient Monte Carlo generation, we apply preselection cuts to
the backgrounds.  Namely, we require at least one jet with $p_T > 30$
GeV and leptons have $p_T > 20$ GeV.  The signal sample is without
preselection requirements.  

\begin{figure}[ht]
\begin{center}
\includegraphics[width=0.4\textwidth]{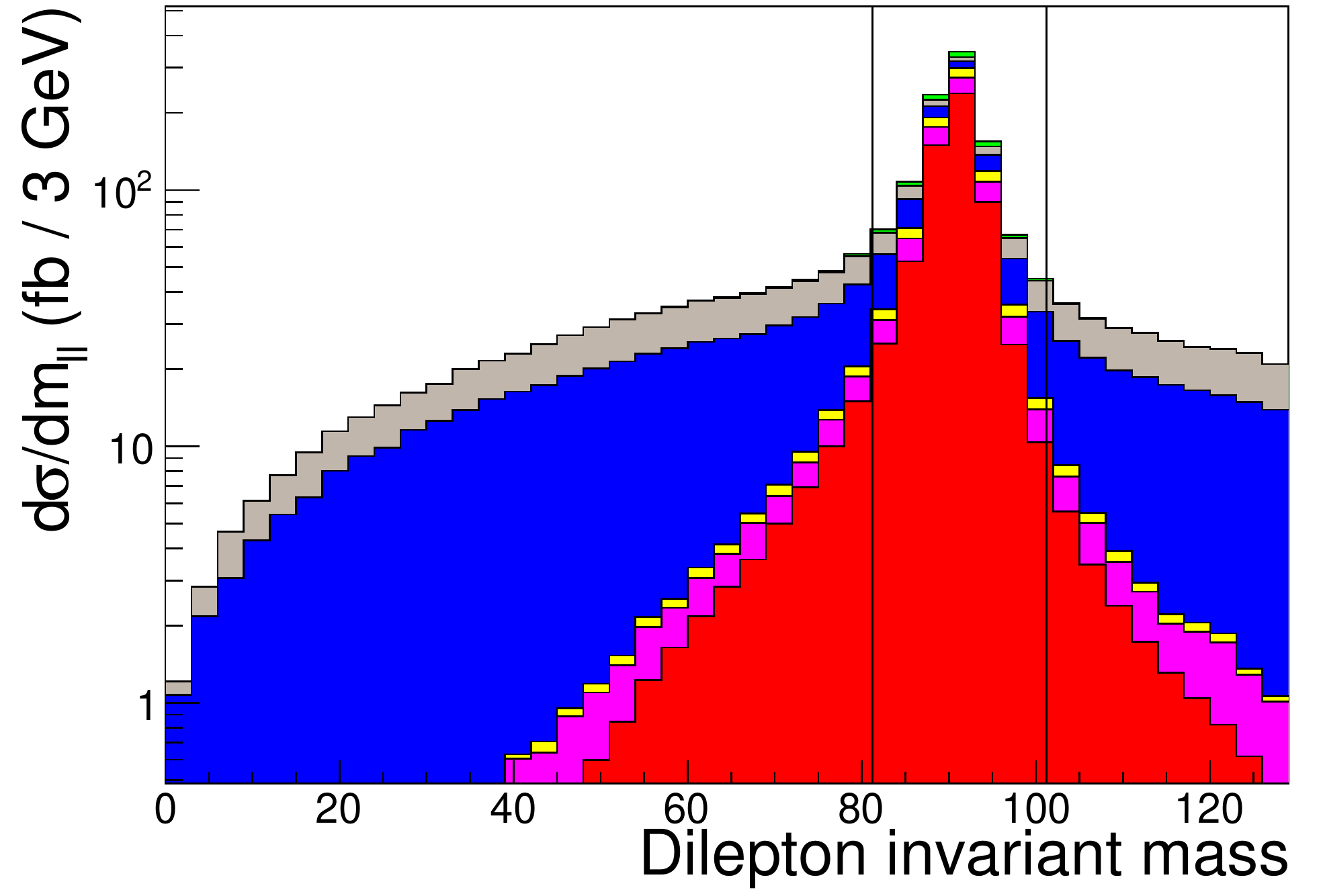}
\caption{Opposite-sign, same flavor lepton pair invariant mass
  distributions for the $h_1 \to \tau^+ \tau^-$ signal benchmark
  (green), and $Z + $ jets $\times 1/500$ (red), $W^\pm Z$ (magenta),
  $ZZ$ (yellow), $t \bar{t} \times 1/10$ (blue), and $W^+ W^-$ (gray)
  backgrounds.  The signal is normalized using $c_{\text{eff}} = 1$.
  The black vertical lines at 81.2 GeV and 101.2 GeV mark the dilepton
  mass window used in our analysis.}
\label{fig:tata_mll}
\end{center}
\end{figure}

Events are clustered with the anti-$k_T$ algorithm using $R = 0.6$,
which will serve as our candidate hadronic di-tau signal jet.  We
select events with at least two leptons with $p_T > 20$ GeV and
$|\eta| < 2.5$, excluding $1.37 < |\eta| < 1.52$ for electron
candidates.  The highest $p_T$ pair of leptons is required to be the
same flavor and opposite sign (OS), and their invariant mass must
fall within 10 GeV of the $Z$ mass.  This helps reduce the
non-resonant dilepton background from $t \bar{t}$ and $W^+ W^-$, as
evident in Fig.~\ref{fig:tata_mll}.

\begin{figure}[ht]
\begin{center}
\includegraphics[width=0.4\textwidth]{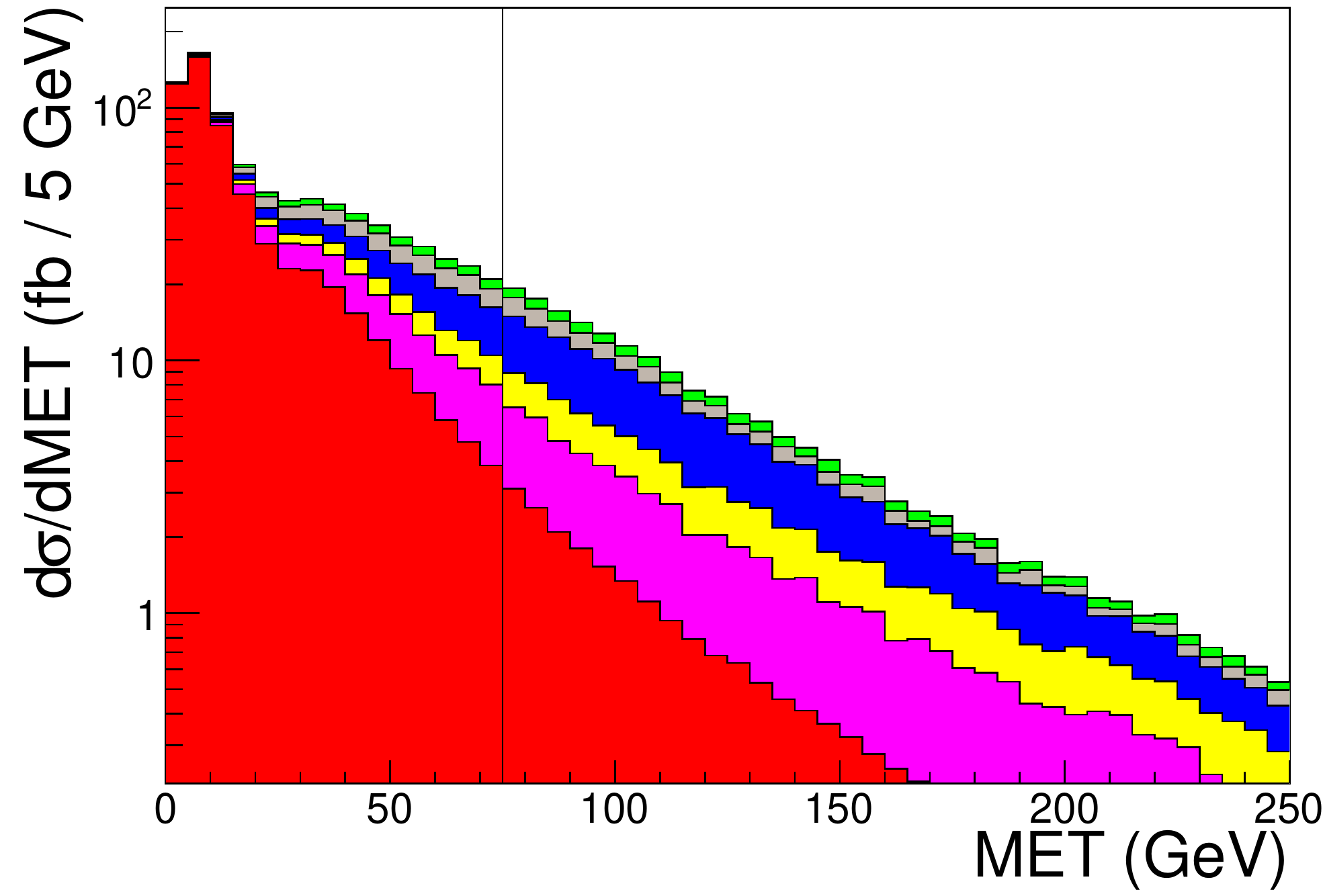}
\caption{MET distributions for the $h_1 \to \tau^+ \tau^-$ signal
  benchmark (green), and $Z + $ jets $ \times \dfrac{1}{500}$ (red),
  $W^\pm Z$ (magenta), $ZZ$ (yellow), $t \bar{t} \times \dfrac{1}{10}$
  (blue), and $W^\pm W^\mp$ (gray) backgrounds.  The signal is
  normalized using $c_{\text{eff}} = 1$.  The black vertical line at
  75 GeV indicates our MET cut.}
\label{fig:tata_MET}
\end{center}
\end{figure}

We next require $\met > 75$ GeV, which helps reduce the $Z + $ jets
background.  Although the $Z + $ jets background has no intrinsic
source for MET, our jet mismeasurement modeling leads to spurious MET
signals.  Also, as opposed to traditional SUSY pair production jets +
MET searches, the mass scale for our hard process is not large, so the
MET tail in our distribution is not highly pronounced, see
Fig.~\ref{fig:tata_MET}.  A future analysis in this channel would
greatly benefit from improving the jet mismeasurement modeling and a
possible subtraction of the $Z + $ jets background from the
differential MET distribution.

\begin{figure}[ht]
\begin{center}
\includegraphics[width=0.4\textwidth]{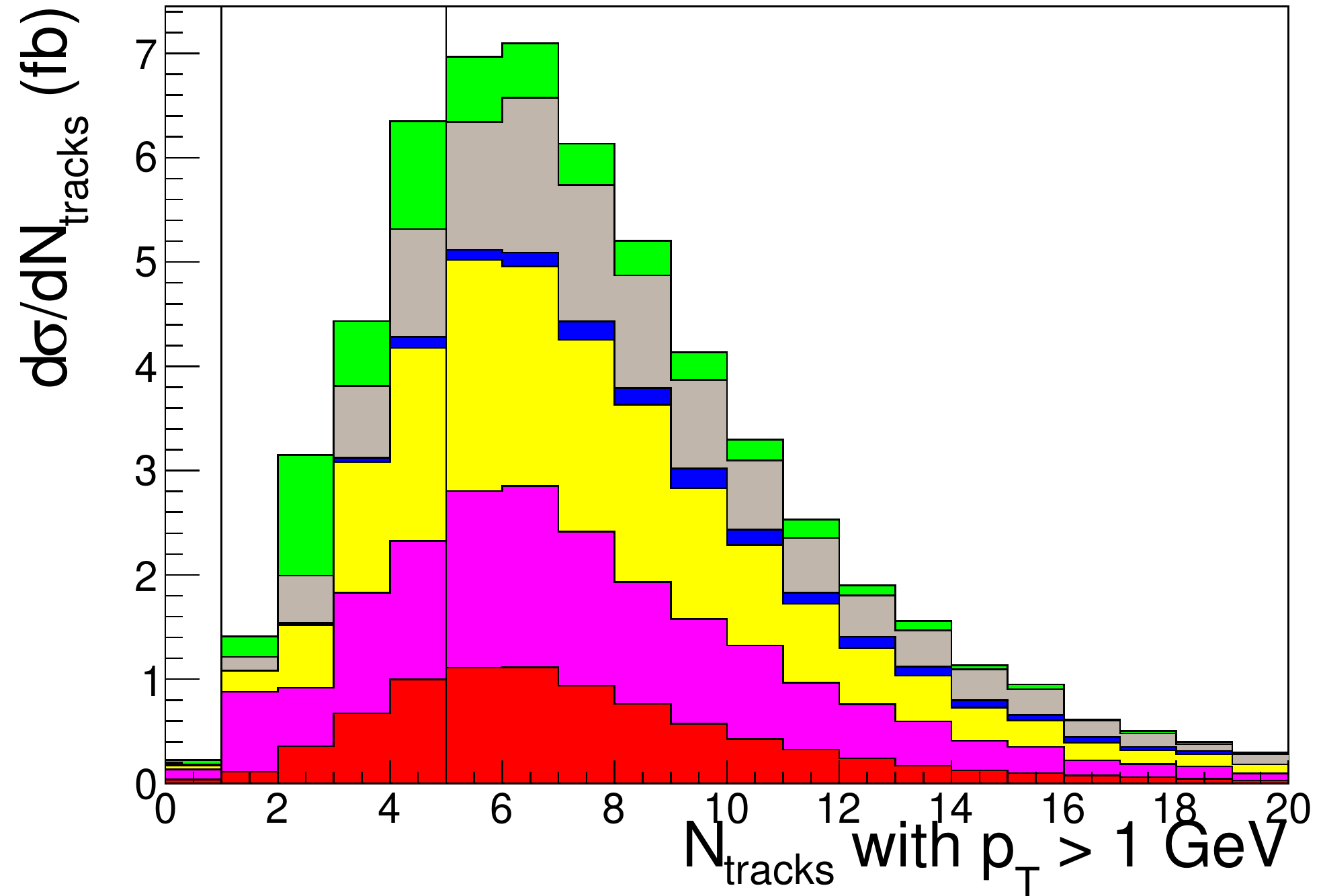}
\caption{Number of charged tracks with track $p_T >1$ GeV for the $h_1
  \to \tau^+ \tau^-$ signal benchmark (green), and $Z + $ jets $
  \times \dfrac{1}{500}$ (red), $W^\pm Z$ (magenta), $ZZ$ (yellow), $t
  \bar{t} \times \dfrac{1}{10}$ (blue), and $W^\pm W^\mp$ (gray)
  backgrounds.  The signal is normalized using $c_{\text{eff}} = 1$.
  We count the number of events with 1 to 5 tracks, indicated by the
  black lines.}
\label{fig:tata_Ntracks}
\end{center}
\end{figure}

We can readily eliminate much of the remaining $t \bar{t}$ background
by vetoing $b$-tagged jets, where we adopt a $b$-tagging efficiency of
60\% for $b$-jets, 10\% for $c$-jets, and a 1\% mistag rate.  We
isolate the $N_{\text{jet}} = 1$ bin and study the track content.
From Fig.~\ref{fig:tata_Ntracks}, we see that the signal
characteristically has fewer hard tracks than the backgrounds, with
prominent peaks at 2 and 4 tracks corresponding to the two one-prong
and three-prong hadronic tau decays.  We sum over 1 to 5 tracks in a
moderately inclusive fashion to help avoid difficulties with track
quality requirements.  The cut efficiencies are listed in
Table~\ref{table:tata}.

\begin{table*}[ht]
\resizebox{\textwidth}{!}{
\begin{tabular}{c|c|c|c|c|c|c}\hline
Cut and      & $Z + h_2$ & $Z + $ jets, $Z \to \ell \ell$ & $t \bar{t} \to b \ell^+ \nu \bar{b} \ell^- \nu$ 
             & $W^+ W^- \to \ell^+ \nu \ell^- \nu$ & $W^\pm Z \to \ell^\pm \nu \ell^+ \ell^-$ & $ZZ \to \ell^+ \ell^- \nu \nu$ \\
Efficiencies & 0.098 $\times c_{\text{eff}}$ pb  
                         & 593.4 pb         & 41.82 pb 
             & 2.412 pb & 0.3461 pb & 0.1299 pb \\
\hline \hline
At least two SF, OS leptons with $p_T > 20$ GeV,
         &        &        &          &          &        &       \\
within $Z$ window
         & 0.4389 & 0.4950 & 3.161E-2 & 3.151E-2 & 0.3113 & 0.4977 \\
MET $ > 75$ GeV
         & 0.1632 & 1.803E-2 & 1.544E-2 & 9.002E-3 & 0.1057 & 0.2269 \\
Require $N_{b-\text{tags}} = 0$, only one jet $p_T > 20$ GeV 
         & 6.052E-2 & 7.046E-3 & 4.03E-4 & 4.729E-3 & 4.226E-2 & 0.1313 \\
Require $1-5$ tracks with $p_T > 3$ GeV
         & 3.710E-2 & 2.739E-3 & 6.526E-5 & 1.464E-3 & 1.575E-2 & 4.712E-2 \\
\hline
Event Number ($500\text{ fb}^{-1}$, $c_{\text{eff}} = 1.0$) 
         & 1800 & 8.10E+5 & 1400 & 1800 & 2700 & 3100 \\
\hline
$S/\sqrt{S+B} (500\text{ fb}^{-1}$, $c_{\text{eff}} = 1.0)$ 
         & \multicolumn{3}{c}{2.0 $\sigma$} \\ 
\hline 
\end{tabular}}
\caption{Cut flow table for $Z h_2$, $Z \to \ell^+ \ell^-$, $h_2 \to
  \chi_1 \chi_1 \tau^+ \tau^-$.  Cross sections for backgrounds
  include preselection cuts of at least one jet with $p_T > 20$ GeV
  and leptons with $p_T > 20$ GeV.  Leptons from decays of gauge
  bosons include $e$, $\mu$, and $\tau$.}
\label{table:tata}
\vspace{1cm}
\end{table*}

After applying all selection cuts, we find the signal can be detected
with $2 \sigma$ expected exclusion sensitivity from 500 fb$^{-1}$ of
LHC 14 TeV data, assuming $c_{\text{eff}} = 1$.  For $c_{\text{eff}} =
0.5$, however, we expect the total HL-LHC luminosity of 3 ab$^{-1}$
would only have $2.4 \sigma$ sensitivity.  This result is mainly
driven by the very large $Z + $ jets background, where the only
effective cuts to reduce this background is our $\met > 75$ GeV
requirement and the track number cut.  As mentioned before, the MET
tail from the $Z + $ jets background arises from our modeling of jet
mismeasurement, which should roughly reproduce the gross features of
an experimental analysis.  But a better understanding of the jet
energy scale will help improve the modeling of the MET tail, likely
leading to improved discovery and exclusion prospects.  An extensive
improvement on the track number requirement is certainly possible, but
would realistically include additional handles to optimize low $p_T$
hadronic tau candidates in a high pile-up environment.  This type of
analysis is beyond the scope of the current work, but we reserve such
a study for future work.

In addition, considering traditional SUSY pair production modes that
lead to $\chi_2$ intermediate states in the cascade would provide a
second source of $h_1 \to \tau^+ \tau^-$ signal events, where the more
energetic kinematics would serve as a stronger handle against the
backgrounds and also lead to improved prospects.  As mentioned
earlier, such production modes could also effectively enable
signal rates to be uncorrelated with the potential improvement in the
invisible and undetected branching fraction of the Higgs, which we
parametrize by $c_\text{eff}$.


\subsection{Case II: $h_2\to b\bar b +\met$}
\label{subsec:bb}

For $m_{h_1} > 10$ GeV, the light Higgs-like scalar dominantly decays
to $b \bar{b}$.  As we are motivated by considering a possible model
for the GC gamma ray excess, we have adopted a benchmark
point consistent with the parameter space scan presented in
Sec.~\ref{subsec:singlino}, which is listed in
Table~\ref{table:benchmark2}.  This also contrasts with our previous
study in Ref.~\cite{Huang:2013ima}, where the $h_1$ mass was much
higher and hence easier to identify from the continuum $Z + $ heavy
flavor jets background.  In this case, the $h_1$ mass is 20 GeV,
leading to a relatively soft $b \bar{b}$ pair.  The event signature
from the $h_2$ cascade is then $b \bar{b} + \met$.  Since the $\met$
signature is only present to the extent that the $b \bar{b}$ system
recoils, we again need a hard object for $h_2$ to recoil against as a
useful trigger and to enhance the MET significance.  For these
purposes, we adopt the $Z h_2$ production mode, with $Z \to \ell^+
\ell^-$, with $\ell = e$ or $\mu$.  The signal is then a dilepton $Z$
candidate, a relatively soft $b$-tagged jet, and $\met$.

\begin{table} [ht]
\begin{tabular}{c|c|c|c|c}\hline\hline
&$m_{h_1}$&$m_{h_2}$& $m_{\chi_1}$ &  $m_{\chi_2}$  \\ \hline
$h_1 \to b \bar b$ & 20 GeV & 125 GeV & 30 GeV & 80 GeV   \\ \hline
\end{tabular}
\caption{Benchmark used for the collider analysis of $h_2 \to b \bar b
  + \met$.}
\label{table:benchmark2}
\end{table}

The main backgrounds for our signal are $Z + $ heavy flavor jets,
including $Z g$, $g \to b \bar{b}$, $g \to c \bar{c}$, and $Zc + Z
\bar{c}$ production, and $t \bar{t}$.  We adopt 60\% for our
$b$-tagging efficiency, 10\% for charm-mistagging, and 1\% for the
remaining light flavor jet mistagging.  While our dominant background
will be from $Zg$ with gluon splitting to two $b$-quarks, there is
still a non-negligible background from the charm-mistag background.

After our mild preselection requirements,~\footnote{We adopt the
  typical default requirements set in MadGraph 5 except for the $Z +
  g$ background, we require the leptonic decay products of the $Z$ to
  have $p_T > 30$ GeV.} the starting cross sections at $\sqrt{s} =
14$ TeV LHC are 48.4 pb for $Z b \bar{b}$, 32.8 pb for $Z c \bar{c}$,
138.9 pb for $Zc + Z\bar{c}$, with subsequent $Z \rightarrow e^+ e^-$,
$\mu^+ \mu^-$, or $\tau^+ \tau^-$ decay, 41.8 pb for $t \bar{t} + $
jets, requiring the fully leptonic decays of the tops, and 0.098
$\times c_{\text{eff}}$ pb for our $Z h_2$ signal, again requiring $Z
\rightarrow e^+ e^-$, $\mu^+ \mu^-$, or $\tau^+ \tau^-$.  The $t
\bar{t}$ and signal cross sections are normalized the same as with the
$\tau^+ \tau^-$ analysis.  Events are clustered with the
angular-ordered Cambridge-Aachen algorithm~\cite{Dokshitzer:1997in,
  Wobisch:1998wt} with distance parameter $R = 1.2$, in order to
capture the larger $b \bar{b}$ system.

\begin{figure}[ht]
\begin{center}
\includegraphics[width=0.4\textwidth]{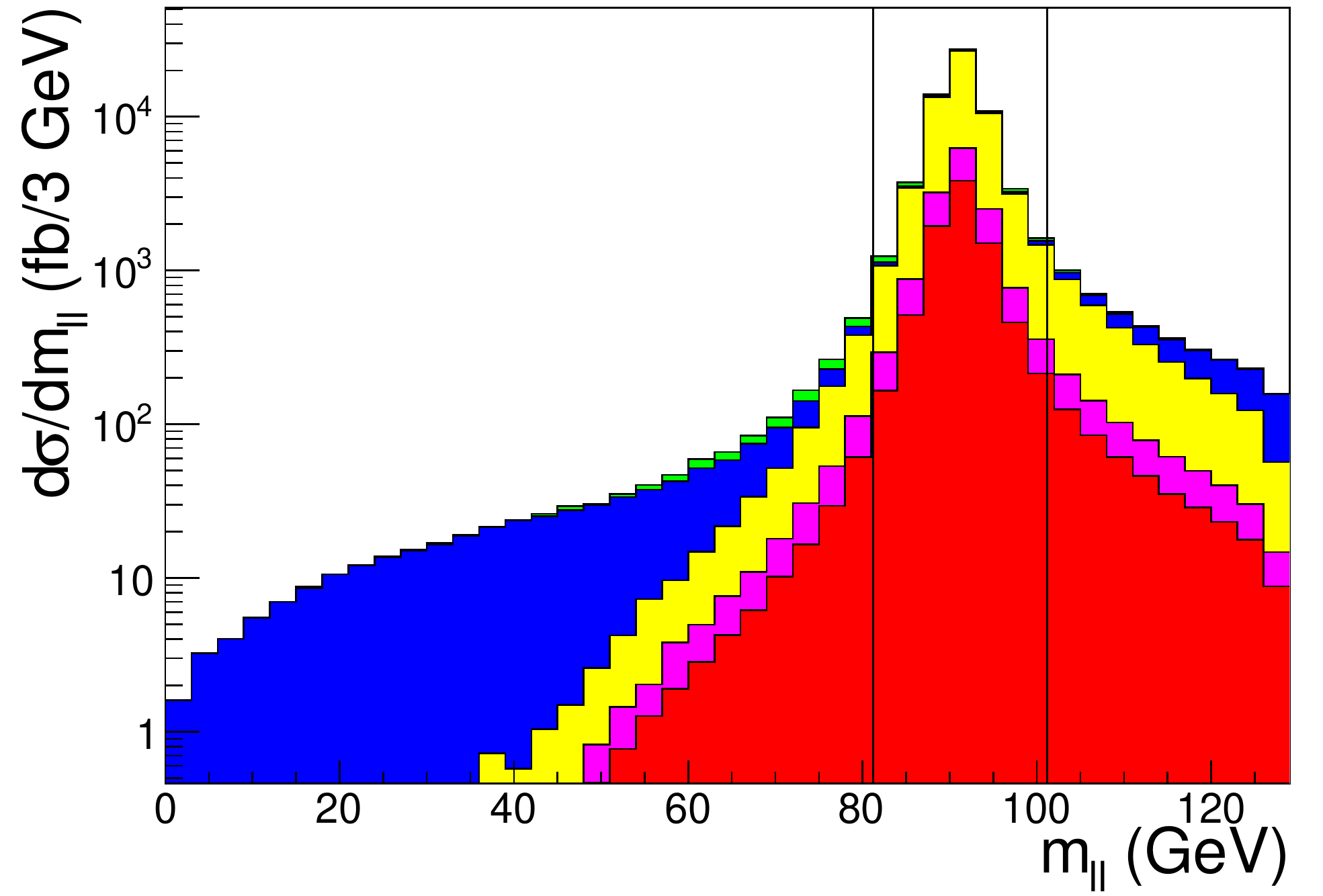}
\caption{Differential cross section vs. $m_{\ell \ell}$ of same
  flavor, opposite sign lepton pairs for signal $\times 200$ (green),
  $Z b \bar{b}$ background (red), $Z c\bar{c}$ background (magenta),
  $Z c, Z \bar{c}$ background (yellow), and $t \bar{t}$ background
  (blue) for the 14 TeV LHC.  The black vertical lines indicate the
  mass window cut used requiring $81.2 < m_{ll} < 101.2$ GeV.  We have
  set $c_{\text{eff}} = 0.5$ for this channel.}
\label{fig:bb_zmass}
\end{center}
\end{figure}

We again start by identifying the leptonic $Z$ candidate, where the
hardest dilepton pair must be an $e^+ e^-$ or $\mu^+ \mu^-$ pair with
each having $p_T > 40$ GeV and satisfying $81.2$ GeV $ < m_{\ell \ell}
< 101.2$ GeV.  Again, this helps to eliminate the non-resonant
dilepton background from $t \bar{t}$, as evident in
Fig.~\ref{fig:bb_zmass}.

\begin{figure}[ht]
\begin{center}
\includegraphics[width=0.4\textwidth]{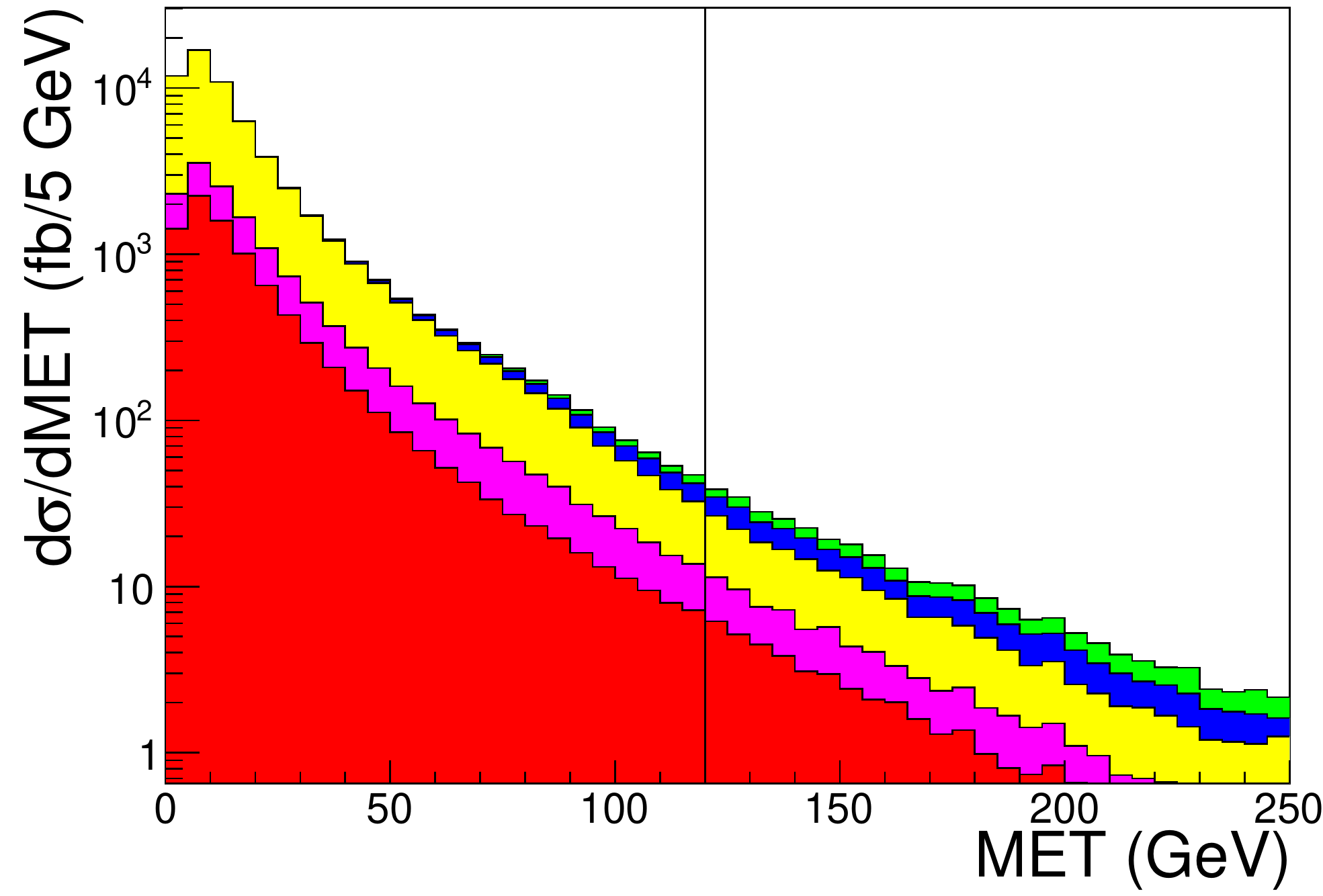}
\caption{Differential cross section vs. MET for signal $\times 20$
  (green), $Z b \bar{b}$ background (red), $Z c\bar{c}$ background
  (magenta), $Z c, Z \bar{c}$ background (yellow), and $t \bar{t}$
  background (blue) for the 14 TeV LHC after the $Z$ mass window cut.
  The black vertical lines indicate our MET $ > 120$ GeV cut.  We have
  set $c_{\text{eff}} = 0.5$. }
\label{fig:bb_MET}
\end{center}
\end{figure}

We then require $\met > 120$ GeV.  In contrast with the $\tau^+
\tau^-$ case, the signal carries a longer and harder $\met$ tail (see
Fig.~\ref{fig:bb_MET}) arising from the different mass splittings in
the benchmark we adopted.  This cut is effective at eliminating the $Z
+ $ heavy flavor jet backgrounds, but the remaining contributions from
these backgrounds are difficult to control because their MET tail
again arises from our jet energy smearing.

\begin{figure}[ht]
\begin{center}
\includegraphics[width=0.4\textwidth]{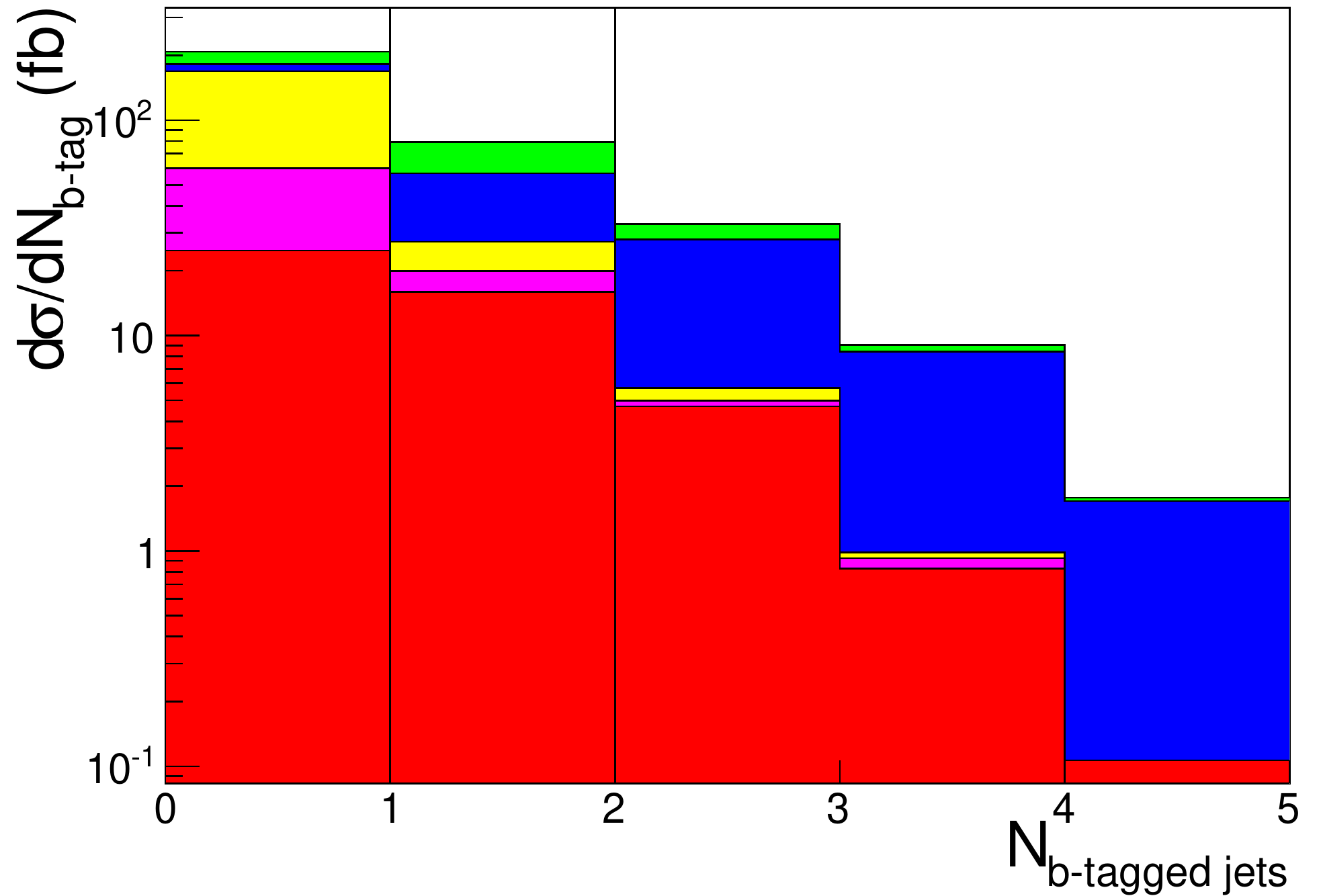}
\caption{Differential cross section vs. $b$-tagged jet multiplicity
  for signal $\times 20$ (green), $Z b \bar{b}$ background (red), $Z
  c\bar{c}$ background (magenta), $Z c, Z \bar{c}$ background
  (yellow), and $t \bar{t}$ background (blue), after the $Z$ mass
  window and MET cuts at the 14 TeV LHC.  The black vertical lines
  indicate the required bin, $N_{b-\text{tags}} = 1$, used in our
  analysis.  We have set $c_{\text{eff}} = 0.5$.}
\label{fig:bb_nbtags}
\end{center}
\end{figure}

We now count the number of $b$-tagged jets with $p_T > 20$ GeV.  Since
the signal is best identified when its two bottom quarks are clustered
into the same jet, as evident in Fig.~\ref{fig:bb_nbtags}, we retain
the 1 $b$-tag bin and discard events with 0 or $\geq 2$ $b$-tagged
jets.

\begin{figure}[ht]
\begin{center}
\includegraphics[width=0.4\textwidth]{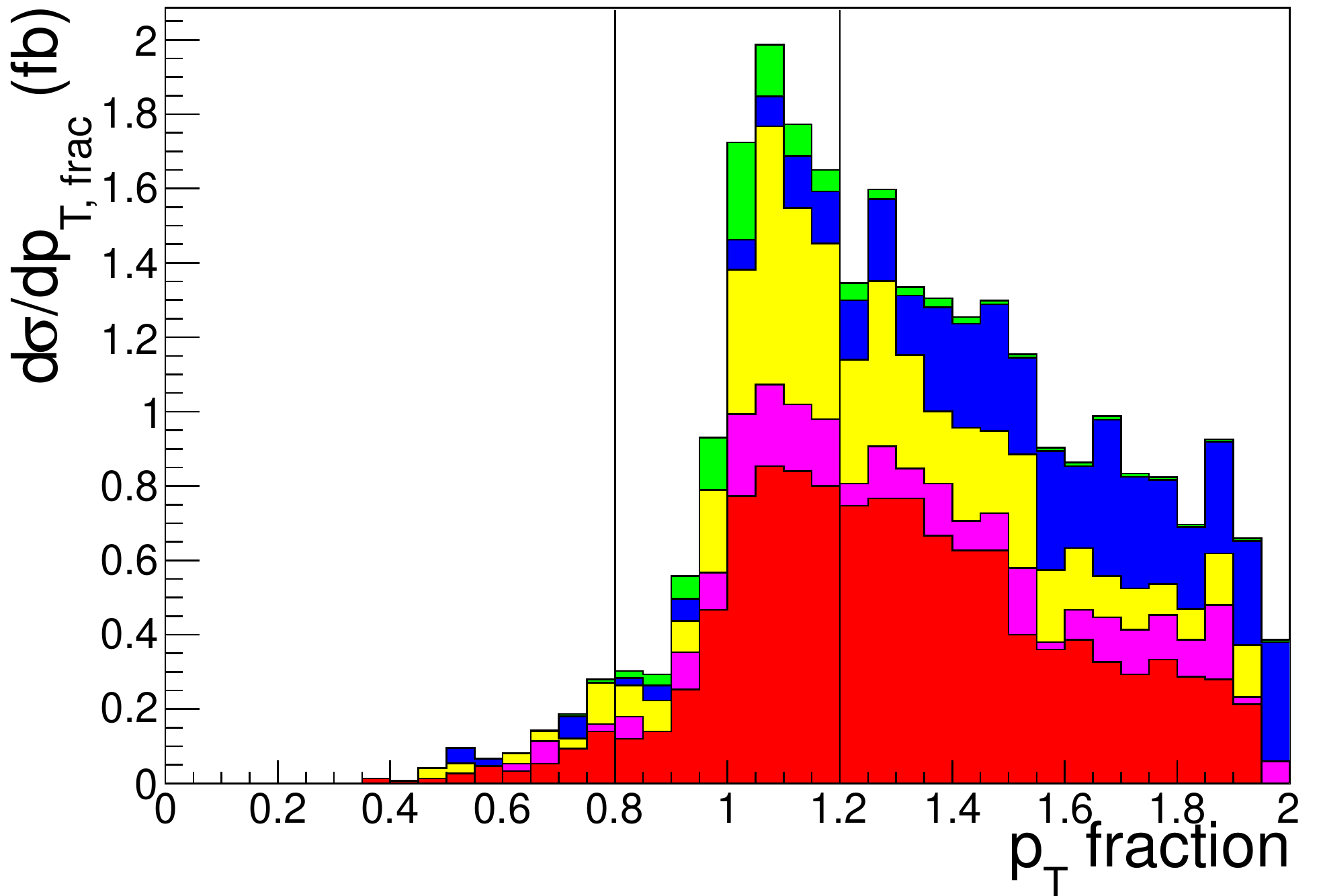}
\caption{Differential cross section vs. $p_{T, {\rm frac}}$, as
  defined in the text, for signal (green), $Z b \bar{b}$ background
  (red), $Z c \bar{c}$ background (magenta), $Z c, Z \bar{b}$
  background (yellow) and $t \bar{t}$ background (blue), after
  applying the $Z$ mass window, MET, and $N_{b-\text{tag}} = 1$ cuts
  at the 14 TeV LHC.  The black vertical lines indicate the $p_{T,
    {\rm frac}}$ window, $0.8 < p_{T, \text{frac}} < 1.2$, requirement
  in our analysis.  We have set $c_{\text{eff}} = 0.5$.}
\label{fig:bb_pTmatch}
\end{center}
\end{figure}

Having isolated the dilepton system as well as the cascade decay of
the $h_2$ boson, we can apply an additional cut with the expectation
that the dilepton system recoils against the collimated $h_2$ cascade
decay.  We construct the scalar sum $p_T$ of the $h_2$ candidate, $p_T
(h_2, \text{ cand}) = p_T (b\text{-jet}) + |\met|$, and then divide it
by the $p_T$ of the dilepton system: $p_{T, \text{ frac}} \equiv p_T
(h_{2, \text{ cand}}) / p_T (\ell \ell_{\text{sys}})$.  Shown in
Fig.~\ref{fig:bb_pTmatch} is the distribution of the transverse
momentum fraction, $p_{T, \text{ frac}} \equiv p_T (h_{2, \text{
    cand}}) / p_T (\ell \ell_{\text{sys}})$.  We observe that the
cutting on $p_{T, \text{ frac}}$ works well at reducing the $t
\bar{t}$ background, where the MET signal tends to arise from
neutrinos of separate decay chains instead of a single cascade decay.
We require $0.8 < p_{T, \text{ frac}} < 1.2$ to isolate well-balanced
$Z h_2$ candidate events.

\begin{table*}[ht]
\resizebox{\textwidth}{!}{
\begin{tabular}{c|c|c|c|c|c}\hline
Cut and      & $Z h_2$ & $Z b \bar{b}$ & $Zc \bar{c}$ 
             & $Zc + Z \bar{c}$ & $t \bar{t}$ \\
Efficiencies & 0.098 $\times c_{\text{eff}}$ pb  
                         & 48.4 pb         & 32.8 pb 
             & 138.9 pb         & 41.8 pb \\
\hline \hline
At least two SF, OS leptons with $p_T > 40$ GeV,
         &        &        &        &        &         \\
within $Z$ window
         & 0.1946 & 0.1774 & 0.1707 & 0.1634 & 0.01193 \\
MET $ > 120$ GeV
         & 5.547E-2 & 9.597E-4 & 1.205E-3 & 4.213E-4 & 1.765E-3 \\
$N_{b-\text{tags}} = 1$, jet $p_T > 20$ GeV 
         & 2.303E-2 & 3.294E-4 & 1.231E-4 & 2.620E-5 & 7.058E-4 \\
$0.8 < p_{T, \text{ frac}} < 1.2$
         & 2.105E-2 & 1.935E-4 & 4.265E-5 & 1.160E-5 & 7.565E-4 \\
\hline
Event Number ($50\text{ fb}^{-1},  c_{\text{eff}} = 0.5$) 
         & 40 & 213 & 53 & 64 & 26 \\
\hline
$S/\sqrt{S+B} (50\text{ fb}^{-1}, c_{\text{eff}} = 0.5)$ 
         & \multicolumn{3}{c}{2.0 $\sigma$} \\ 
\hline 
\end{tabular}}
\caption{Cut flow: Analysis cuts and efficiency table for the $h_1
  \rightarrow b \bar{b}$ channel, $m_{h_1} = 20$, $\chi_1 = 30$,
  $\chi_2 = 80$ GeV.  The decays $Z \to \ell^+ \ell^-$ and $W \to \ell
  \nu$, $\ell = e$, $\mu$, $\tau$ are included in the quoted cross
  sections.}
\vspace{1cm}
\label{table:bb}
\end{table*}

After these cuts, we find that we have 2.0$\sigma$ exclusion
sensitivity with 50 fb$^{-1}$ of 14 TeV LHC luminosity.  For
$c_{\text{eff}} = 0.1$, which is the expectation for the ultimate
sensitivity of the LHC to an exotic decay mode of the SM-like Higgs
via coupling fits~\cite{Curtin:2013fra}, we estimate that about 1.2
ab$^{-1}$ of 14 TeV LHC luminosity is required for 2.0$\sigma$
sensitivity, although this luminosity scaling does not include any
estimation of systematic effects.  Complete cut flow information and
sensitivity calculation are presented in Table~\ref{table:bb}.  The
dominant background is from $Z b \bar{b}$, which could be further
reduced with a grooming procedure looking for hard subjets if the
signal were further on the $b \bar{b}$ continuum tail, as demonstrated
in~\cite{Huang:2013ima}.  For our current benchmark, however, the
lighter $h_1$ mass and the softer subjets render the grooming
procedure ineffective at resolving the signal from the continuum
background.  Alternative jet substructure techniques, however, could
be promising tools to resolve the $h_1 \to b \bar{b}$ signal bump, but
additional handles on issues like subjet resolution and pile-up
mitigation would also need to be taken into account.  As with the $h_1
\to \tau^+ \tau^-$ scenario, further improvements could be made by
studying traditional pair production modes of supersymmetric
particles.  The possibility of identifying single jets with multiple
displaced vertices, corresponding to multiple $b$-hadron candidates,
would also be a promising avenue for signal extraction.  We will leave
these interesting questions for future work.



\section{Conclusions}
\label{sec:conclusion}

In this article, we emphasized that the nearly PQ-symmetry limit
provides a supersymmetric benchmark for both a singlino-like sub-EW
scale DM, as well as novel exotic decays of SM-like Higgs, $h_2 \to
\chi_1 \chi_2$ and $h_2 \to \chi_2 \chi_2$, with the bino-like
$\chi_2$ further decaying in multiple ways.  The collider signature 
of this new category of exotic Higgs decays is characterized by MET and
some visible objects, with or without a resonance. 
We have pursued the $h_2 \to \chi_1 \chi_2 \to h_1/a_1 \chi_1 \chi_1 $ mode,
which is favored more by kinematics, and presented
analyses of the $\tau^+ \tau^-$ and $b \bar{b}$ channels for
extracting these signals from the SM backgrounds.

In the first analysis, we studied a benchmark model with $m_{h_1} = 8$
GeV, $m_{\chi_1} = 10$ GeV, and the decay $h_2 \to \tau^+ \tau^-
\met$, where we adopted a track-based identification of pair of
hadronic taus decays.  We motivated the $Z h_2$, $Z \to \ell^+ \ell^-$
production mode as the easiest trigger path, but nevertheless the
relatively moderate MET from our signal implies the very large
background from $Z + $ jets cannot be dramatically reduced.  We
estimate that roughly 500 fb$^{-1}$ of 14 TeV LHC luminosity is
required to have $2 \sigma$ exclusion sensitivity to this channel.
Improving this channel would require new kinematic handles or
alternative $\tau$ decay channels.

In the second analysis, we studied a GC gamma ray excess
benchmark with $m_{h_1} = 20$ GeV, $m_{\chi_1} = 30$ GeV and $h_1
\rightarrow b \bar{b}$.  The $Z h_2$, $Z \to \ell^+ \ell^-$ production
mode again provided a good trigger path, and the effective $b$-tagging
requirements and larger MET signature of our signal were significantly
more effective at reducing SM backgrounds.  Using $c_{\text{eff}} =
0.5$, we estimate that 50 fb$^{-1}$ of 14 TeV LHC luminosity is needed
to reach $2 \sigma$ exclusion sensitivity for this benchmark.  As
reiterated from~\cite{Huang:2013ima}, one novel cut used in this
analysis took advantage of the fact that for the decay topology in
Fig.~\ref{fig:decay_topo}, the MET arises only from the $h_2$ decay, in
contrast to the usual MET signatures of pair-produced MSSM
superpartners or SM $t \bar{t}$ background.

The possibilities other than the ones discussed here ({\it e.g.}, $h_2
\to \chi_2 \chi_1$ with different $\chi_2$ decay modes, as well as
$h_2 \to \chi_2 \chi_2$) can lead to collider signatures and
kinematics requiring different analyses from those we have presented.
Additional production modes, such as those arising from pair
production of superpartners, are also promising probes for studying
the singlet-like $h_1$ and singlino-like $\chi_1$ states.  We will
leave these interesting topics for a future study.


{\bf [Note added]}: While this article was in preparation, the papers
by T. Han, {\it et. al.}~\cite{Han:2014nba} and by C. Cheung, {\it
  et. al.}~\cite{Cheung:2014lqa} appeared, which partially overlap
with this one in discussing the potential role of the nearly
PQ-symmetry limit of the NMSSM, a supersymmetric benchmark for sub-EW
scale singlino-like DM~\cite{Draper:2010ew}, in explaining the GC
gamma ray excess.  A notable difference, however, is that we emphasize
the connection between sub-EW scale DM and the exploration of
semi-visible exotic decays of the 125 GeV Higgs boson, MET + visible,
at colliders~\cite{Huang:2013ima}, while~\cite{Han:2014nba} is focused
on the mechanisms to achieve correct relic density
and~\cite{Cheung:2014lqa} is dedicated to the study on explaining the
GC gamma-ray excess in supersymmetric scenarios.

\section*{Acknowledgments}

We would like to thank Brock Tweedie, Patrick Draper, Michael
Graesser, Joe Lykken, Adam Martin, Nausheen Shah, Jessie Shelton, Matt
Strassler, and Carlos Wagner for useful discussions.  TL is supported
by his start-up fund at the Hong Kong University of Science and
Technology.  JH is supported by the DOE Office of Science and the LANL
LDRD program.  JH would also like to thank the hospitality of
University of Washington, where part of the work was finished.  L-TW
is supported by the DOE Early Career Award under Grant DE-SC0003930.
L-TW is also supported in part by the Kavli Institute for Cosmological
Physics at the University of Chicago through NSF Grant PHY-1125897 and
an endowment from the Kavli Foundation and its founder Fred Kavli.  FY
would like to thank the Theoretical High Energy Physics group at
Johannes Gutenberg Universit\"{a}t Mainz for their hospitality, where
part of this work was completed.  Fermilab is operated by the Fermi
Research Alliance, LLC under Contract No. DE-AC02-07CH11359 with the
US Department of Energy.

We also would like to acknowledge the hospitality of the Kavli
Institute for Theoretical Physics and the Aspen Center for Physics,
where part of this work was completed, and this research is supported
in part by the National Science Foundation under Grant No. NSF
PHY11-25915.

\vspace*{-15pt}

\appendix

\section{Mass Eigenvalues and Eigenstates of the $CP$-even 
Higgs Sector in the PQ Limit}
\label{app:gauge}

The mass eigenvalues of the $CP$-even Higgs sector in the Peccei-Quinn
limit are given by
\begin{eqnarray}
m_{h_1}^2  &=&  - \frac{4(\lambda^2 v^2 \mu^2 \varepsilon'^2)}{m_Z^2} 
\nonumber \\ 
&+& \frac{4 \lambda^2 v^2  }{m_Z^6}  \Big(4 v^2 \varepsilon'^4 \lambda^2 \mu^4  
+ \frac{m_Z^4 (1 - \varepsilon') (m_Z^2 + 2 \varepsilon' \mu^2)}{\tan^2 \beta} 
\Big) \nonumber \\
&+& \sum\limits_i \mathcal{O} \left( \frac{ \lambda^{5-i}}{\tan^i \beta} 
\right) \ , \nonumber \\
\vspace{0.1cm}
m_{h_2}^2  &=&  m_Z^2 + \left(\frac{-4 m_Z^2}{\tan^2\beta} 
+ \frac{4 v^2 \varepsilon'^2 \lambda^2 \mu^2}{m_Z^2}\right) 
+ \sum\limits_i \mathcal{O} \left( \frac{\lambda^{3-i}}{\tan^i \beta} 
\right) \ , \nonumber \\
\vspace{0.1cm}
m_{h_3}^2  &=& (1 + \varepsilon') \mu^2 \tan^2 \beta + (1 + \varepsilon') 
\mu^2 \nonumber \\ 
&+& \left( \frac{3 m_Z^2}{\tan^2 \beta} + v^2 (1 + \varepsilon') 
\lambda^2 \right) +
\sum\limits_i \mathcal{O} \left( \frac{\lambda^{3-i}}{\tan^i \beta} 
\right) \ .
\label{eigenmass}
\end{eqnarray}
In the extremal limit $\lambda = 0$, $m_{h_2}^2$ is reduced to 
\begin{eqnarray}
m_{h_2}^2  &=&  m_Z^2 + \frac{-4 m_Z^2}{\tan^2 \beta} +
\sum\limits_i \mathcal{O} \left( 
\frac{\lambda^{3-i}}{\tan^i \beta} \right) \nonumber \\
&=& M_Z^2\cos^2 2\beta + 
\sum\limits_i \mathcal{O} \left( 
\frac{\lambda^{3-i}}{\tan^i \beta} \right)
\end{eqnarray}
a familiar result in the MSSM. The eigenstates of the three $CP$-even
Higgs are given by
\begin{eqnarray}
S_{1d} &=& 
\frac {\lambda v}{\tan \beta} 
\left( \frac{1}{\mu} + \frac{2 \varepsilon' \mu}{m_Z^2}\right) + 
\sum\limits_i \mathcal{O} 
\left( \frac{\lambda^{3-i}}{\tan^i \beta} \right) \ , \nonumber \\
S_{1u} &=& \frac{2 v \varepsilon' \lambda \mu}{m_Z^2} + 
\sum\limits_i \mathcal{O} 
\left( \frac{\lambda^{3-i}}{\tan^i \beta} \right) \ , \nonumber \\
S_{1s} &=& 1 + 
\sum\limits_i \mathcal{O} 
\left( \frac{\lambda^{3-i}}{\tan^i \beta} \right) \ , \nonumber \\
S_{2d} &=& \frac{1}{\tan \beta} + 
\sum\limits_i \mathcal{O} 
\left( \frac{\lambda^{2-i}}{\tan^i \beta} \right) \ , \nonumber \\
S_{2u} &=& 1 +
\sum\limits_i \mathcal{O} 
\left( \frac{\lambda^{2-i}}{\tan^i \beta} \right) \ , \nonumber \\
S_{2s} &=& -\frac{ 2\lambda \varepsilon' v \mu}{m_Z^2} + 
\sum\limits_i \mathcal{O} 
\left( \frac{\lambda^{2-i}}{\tan^i \beta} \right) \ , \nonumber \\
S_{3d} &=& 1 +
\sum\limits_i \mathcal{O} 
\left( \frac{\lambda^{2-i}}{\tan^i \beta}\right) \ , \nonumber \\
S_{3u} &=& -\frac{1}{\tan \beta} +
\sum\limits_i \mathcal{O} 
\left( \frac{\lambda^{2-i}}{\tan^i \beta} \right) \ , \nonumber \\
S_{3s} &=& 0 + 
\sum\limits_i \mathcal{O} 
\left( \frac{\lambda^{2-i}}{\tan^i \beta} \right) \ .
\label{eqn:eigenstate}
\end{eqnarray}
For our purposes, the eigenvalue and eigenstate of the lightest
$CP$-even Higgs boson are calculated to an order above the other two.

\section{Calculation of $y_{h_2 a_1 a_1}$}
\label{app:goldstone}

In this Appendix, we calculate the coupling $y_{h_2 a_1 a_1}$ using
the properties of the Goldstone boson.  In the polar coordinates of
the Higgs fields, $y_{h_2 a_1 a_1}$ arises from the kinetic term of
$a_1$.  We will take the kinetic term of $H_d$ as an illustration.  We
write $H_d$ as
\begin{eqnarray}
H_d &=& \left(v_d + \frac{S_{1d} h_1 + S_{2d} h_2 +S_{3d} h_3}{\sqrt{2}} 
\right) \nonumber \\
&& \times \exp \left( \frac{i (P_{1d} a_1 + P_{2d} a_2 +P_{3d} a_3)}{v_d} 
\right) \ .
\end{eqnarray} 
The relevant term in the
$H_d$ kinetic term expansion is
\begin{eqnarray}
\partial H_d^* \partial H_d \sim  \sqrt{2} \frac{S_{2d}P_{1d}^2} {v_d} 
h_2 \partial a_1 \partial a_1 \ .
\end{eqnarray}
Because $a_1$ is massless, we have $\partial a_1 \partial a_1 =
p_{a_1}^2 a_1^2 = \frac{m_{h_2}^2}{2}=\frac{m_{Z}^2}{2} +
\sum\limits_i \mathcal{O} \left( \lambda^{2-i} / \tan^i \beta
\right)$. So the contribution of the $H_d$ kinetic term to the $y_{h_2
  a_1 a_1}$ is
\begin{equation}
\frac{S_{2d} m_{h_2}^2 P_{1d}^2}{\sqrt{2} v_d} \ .
\end{equation}
With all Higgs kinetic terms incorporated, we have
\begin{eqnarray}
y_{h_2 a_1 a_1} &=& \frac{m_Z^2}{\sqrt 2} \Big( 
\frac{S_{2d} m_{h_2}^2 P_{1d}^2}{v_d} +  
\frac{S_{2u} m_{h_2}^2 P_{1u}^2}{v_u}  \nonumber \\
&+& \frac{S_{2s} m_{h_2}^2 P_{1s}^2}{v_S} \Big ).
\end{eqnarray}
with $P_{1d,1u,1s}$ defined in
Eqs.~(\ref{eqn:axion1})$-$(\ref{eqn:axion3}) after decomposing the
superfields and isolating into pseudoscalar components.  This
immediately reproduces Eq.~(\ref{eqn:yh2a1a1}).

\section{Further discussion of the saxion mass}
Unlike the axino, the saxion may obtain sizable mass corrections in
its diagonal mass term or via its mixing with other massive particles,
with SUSY softly broken.  The diagonal saxion mass at tree level is
given by
\begin{equation}
m_s^2 = \sum_{i,j} \frac{q_i v_i }{v_{\rm PQ}} M_{ij} 
\frac{q_j v_j }{v_{\rm PQ}}  
\end{equation}
with 
\begin{equation}
M_{ij} = \left.
\frac{\partial^2 V}{\partial \phi_i   \partial \phi_j} + 
\frac{\partial^2 V}{\partial \phi_i^* \partial \phi_j} 
\right|_{\phi_{i,j}, \phi_{i,j}^*  = v_{i,j}}
\end{equation}
being the squared mass matrix.  Then we see
\begin{equation}
m_s^2 = 4 v^2 (1 + \varepsilon') \lambda^2 \ , \text{ with } 
\varepsilon' = \frac{A_\lambda}{\mu \tan \beta} -1 \ ,
\end{equation}
where $A_\lambda$ is the soft trilinear scalar coupling of
Eq.~(\ref{eqn:PQlimit}).  The corrections from mixing, however, can be
of the same order.  With the mixing corrections included, the
tree-level saxion mass is given by
\begin{eqnarray}
m_s^2 &=& - \frac{4 (\lambda^2 v^2 \mu^2 \varepsilon'^2)}{m_Z^2} 
          \nonumber \\
&+& \frac{4 \lambda^2 v^2  }{m_Z^6}  
\Big(4 v^2 \varepsilon'^4 \lambda^2 \mu^4  + 
\frac{m_Z^4 (1 - \varepsilon') (m_Z^2 + 2 \varepsilon' 
\mu^2)}{\tan^2\beta} \Big) \nonumber \\
&+& \sum\limits_i \mathcal{O} \left( 
\frac{\lambda^{5-i}}{\tan^i \beta} \right) \ .
\end{eqnarray}
At tree level, the first term is negative, while the second term is
smaller than the first by a factor $\sum\limits_i \mathcal{O} \left(
\frac{\lambda^{2-i}}{\tan^i \beta} \right)$.  Thus, to avoid a
tachyonic vacuum,
\begin{equation}
\varepsilon'^2 < \frac{m_Z^2}{\mu^2 \tan^2\beta}
\label{eqn:stability}
\end{equation}
is required~\cite{Draper:2010ew}.  This condition sets an upper bound
for the tree-level $m_s^2$
\begin{equation}
m_s^2 < \frac{4 v^2 \lambda^2}{\tan^2 \beta} 
\end{equation}
which is $(\mathcal{O}(10) \text{ GeV})^2$ or lower in the context we
consider.

The $\mathcal{O}(10)$ GeV saxion mass feature can be seen in another
way.  It is easy to calculate the determinant of the mass squared
matrix of the $CP$-even Higgs bosons, which is given by
\begin{eqnarray}
\det M &=& -4 \lambda^2 v^2 \tan \beta^2 \mu^4 
(1 + \varepsilon') \nonumber \\
&& \times \left (\varepsilon'^2 - \frac{m_Z^2 (1 - \varepsilon') - 
\varepsilon' (2 + \varepsilon') \mu^2 }{\mu^2 \tan^2\beta} \right) 
\nonumber \\
&& + \sum\limits_i \mathcal{O} 
\left( \frac{\lambda^{4-i}}{\tan^i \beta} \right) \ .
\end{eqnarray}
To avoid a tachyonic vacuum at tree-level, the determinant must be
positive, which immediately leads to the condition given in
Eq.~(\ref{eqn:stability}) if $\varepsilon' > -1$.

We now address the question of why the saxion mass does not get a
large mass correction from the soft SUSY breaking mass term of the
singlet field, even though the saxion is singlet-like.  Recall the
vacuum stability conditions are twofold.  First, the Higgs potential
must be locally flat at the vacuum point, so its first-order partial
derivative with respect to the field variables must be zero.  Second,
the physical masses of the scalar particles cannot be negative, which
implies that the second-order partial derivative with respect to the
mass eigenstates must be positive.  The latter has been discussed
above via the determinant argument.  To see what the former implies,
we start by expressing the scalars as
\begin{eqnarray}
H_u^0 &=& v_u + \frac{H_{u}^R + iH_{u}^I}{\sqrt{2}} \,,\quad \nonumber \\
H_d^0 &=& v_d + \frac{H_{d}^R + iH_{d}^I}{\sqrt{2}} \,,\quad \nonumber \\
S &=& v_S + \frac{S^R + iS^I}{\sqrt{2}} \ .
\label{eqn:204}
\end{eqnarray}
We can derive the locally-flat conditions for the $CP$-even Higgs
components~\footnote{In this article, we assume that there is neither
  explicit nor spontaneous $CP$-violation. In this case, the
  locally-flat conditions with respect to $CP$-odd Higgs components
  are satisfied automatically.} and re-express the soft SUSY-breaking
Higgs masses in terms of $\lambda$, $A_\lambda$, $v_u$, $v_d$ and
$v_S$.  At tree level, they are given by
\begin{eqnarray}
m_{H_d}^2 &=& -\mu^2 + B_\mu \tan \beta - \frac{m_Z^2}{2} 
\cos 2\beta \ , \nonumber \\
m_{H_u}^2 &=& -\mu^2 + B_\mu \cot \beta + \frac{m_Z^2}{2} 
\cos 2\beta \ , \nonumber  \\
m_S^2  &=& -\lambda^2 v^2 + A_{\lambda} \lambda^2 v^2 
\frac{\sin 2\beta}{2 \mu} \ ,
\label{eqn:104}
\end{eqnarray}
where
\begin{equation}
B_\mu = \mu A_{\lambda} - \frac{\lambda^2 v^2
\sin 2\beta}{2} \ ,
\end{equation}
and $\mu = \lambda v_S$ as usual.  For $\tan^2 \beta \gg 1$, the right
hand side of Eq.~(\ref{eqn:104}) can be expanded as
\begin{eqnarray}
m_{H_d}^2 &=& \mu^2 \tan^2 \beta (1 + \varepsilon') \nonumber \\
&& -\mu^2 - \lambda^2 v^2 + \frac{M_Z^2}{2} + 
\mathcal{O} \left( \frac{1}{\tan^2 \beta} \right) \ , 
\label{eqn:303} \\
m_{H_u}^2 &=& \mu^2 \varepsilon' - \frac{M_Z^2}{2} + 
\mathcal{O} \left( \frac{1}{\tan^2 \beta} \right) \ , 
\label{eqn:304} \\
m_S^2  &=& \lambda^2 v^2 \varepsilon' + 
\mathcal{O} \left( \frac{1}{\tan^2 \beta} \right) \ . 
\label{eqn:211}
\end{eqnarray}
So, $m_S^2$ needs to be much smaller than the squared EW scale to get
a stable vacuum.  Thus, in the PQ symmetry limit, the vacuum stability
forbids the saxion to obtain a large mass correction from softly SUSY
breaking effects.


\end{document}